\documentclass[final]{elsarticle}
\usepackage{hyperref}
\usepackage{geometry}
\biboptions{sort&compress}
\def\snm#1{\textcolor{black}{#1}}  %
\usepackage{framed,multirow}
\usepackage[utf8]{inputenc}
\usepackage[T1]{fontenc}
\usepackage{amssymb,amsmath}
\usepackage{latexsym}
\usepackage{algorithmicx,algorithm,algpseudocode}
\usepackage{url}
\usepackage{xcolor}
\definecolor{newcolor}{rgb}{.8,.349,.1}
\usepackage{framed}
\journal{Journal of computational physics}

\usepackage{siunitx}
\usepackage[center]{subfigure}
\newcommand{\ud}{\, \mathrm{d}}	%
\providecommand{\abs}[1]{\left\lvert #1\right\rvert}

\renewcommand{\i}{\mathrm{i}}
\renewcommand{\j}{\mathrm{j}}
\newcommand{\e}{\mathrm{e}}
\newcommand{\nuv}{\boldsymbol{\nu}}
\newcommand{\alphav}{\boldsymbol{\alpha}}
\newcommand{\Lv}{\mathbf{L}}
\newcommand{\xv}{\mathbf{x}}
\newcommand{\qv}{\mathbf{q}}

\newcommand{\sv}{\mathbf{s}}
\newcommand{\Sv}{\mathbf{S}}
\newcommand{\vv}{\mathbf{v}}
\newcommand{\nv}{\mathbf{n}}
\newcommand{\Fv}{\mathbf{F}}
\newcommand{\Kv}{\mathbf{K}}
\newcommand{\Mv}{\mathbf{M}}

\newcommand{\kv}{\mathbf{k}}

\newcommand{\betav}{\boldsymbol\beta}

\newcommand{\pair}[2]{\lbrace#1,\,#2\rbrace}

\newcommand{\Q}{\mathcal{Q}}

\newcommand{\EP}[1]{EP$_#1$}
\graphicspath{{Figures/}{.}}

\makeatletter
\def\ps@pprintTitle{%
 \def\@oddfoot{\hfill\@date}%
 }
\makeatother

\begin{document}
\begin{frontmatter}

	\title{Fast recovery of parametric eigenvalues depending on several parameters and location of high order exceptional points}
	\author[1]{Benoit \snm{Nennig}\corref{cor1}}
	\ead{benoit.nennig@isae-supmeca.fr}
	\cortext[cor1]{Corresponding author}
	
	\author[1]{Martin \snm{Ghienne}}
	\ead{Martin.Ghienne@isae-supmeca.fr}

	\author[2]{Emmanuel \snm{Perrey-Debain}}
\ead{Emmanuel.Perrey-Debain@utc.fr}	
	\address[1]{Institut sup\'erieur de m\'ecanique de Paris (ISAE-Supméca), Laboratoire Quartz EA 7393,\\
	3 rue Fernand Hainaut, 93407 Saint-Ouen, France.}
	\address[2]{Université de technologie de Compiègne, Roberval (Mechanics, energy and electricity), Centre de recherche Royallieu, CS 60319 - 60203, Compiègne Cedex, France}

	\date{\today}

	\begin{abstract}
A numerical algorithm is proposed to deal with parametric eigenvalue problems involving non-Hermitian matrices and is  exploited to find location of defective eigenvalues in the parameter space of non-Hermitian parametric eigenvalue problems.
These non-Hermitian degeneracies also called exceptional points (EP) have raised considerable attention in the scientific community as these can have a great impact in a variety of physical problems. 
The method first requires the computation of high order derivatives of a few selected eigenvalues with respect to each parameter involved. The second step is to recombine these quantities to form new coefficients associated with a partial characteristic polynomial (PCP). By construction, these coefficients are regular functions in a large domain of the parameter space which means that the PCP allows one to recover the selected eigenvalues as well as the localization of high order EPs by simply using standard root-finding algorithms.

The versatility of the proposed approach is tested on several applications, from mass-spring systems to guided acoustic waves with absorbing walls and room acoustics. The scalability of the method to large sparse matrices arising from conventional discretization techniques such as the finite element method is demonstrated.

The proposed approach can be extended to a large number of applications where EPs play an important role in quantum mechanics, optics and photonics or in mechanical engineering.
	\end{abstract}

	\begin{keyword}
	Parametric eigenvalue problem \sep Veering \sep Exceptional point \sep  Puiseux series \sep Defective eigenvalue \sep Singular perturbation \sep Stability analysis
	\end{keyword}

\end{frontmatter}

\setcounter{tocdepth}{4}
\setcounter{secnumdepth}{4}

\section{Introduction}

In this paper, we are concerned with the numerical treatment of parametric eigenvalue problems
which take the general form
\begin{equation}\label{eq:PVP}
\Lv (\lambda(\nuv), \nuv) \xv (\nuv) =\mathbf{0},
\end{equation}
where, for a given vector $\nuv =(\nu_1,\nu_2,\ldots,\nu_N) \in \mathbb{C}^N$ which contains $N$ independent complex-valued parameters, $\lambda(\nuv)$ is an eigenvalue and  $\xv(\nuv)\neq \mathbf{0}\in \mathbb{C}^M$ is the associated right eigenvector. Here the matrix function $\Lv\in \mathbb{C}^{M\times M}$ admits the decomposition
\begin{equation}%
\Lv (\lambda, \nuv)  =\sum_{j \geq 0} f_j(\lambda) \Kv_j(\nuv),
\end{equation}
where $f_j$ is a polynomial function and matrices $\Kv_j$
are supposed to be analytic functions of the parameters vector $\nuv$. The decomposition encompasses most situations encountered in the scientific literature and $\Lv$ may represent for instance a standard, a generalized or a general quadratic eigenvalue problem.
The aim of this paper is to propose a scalable algorithm in order to explore efficiently the trajectories of a given subset of eigenvalues in a large domain of the parameter space.  The efficiency of the method relies on the construction of a partial characteristic polynomial based on the Taylor expansion of each eigenvalue of the subset. The work presented in the paper is twofold: first, it  develops a novel method for solving eigenvalue perturbation problems; second, it applies the method to the computation of exceptional points (EPs).
EPs are special degeneracies corresponding to specific values of the system parameters for which both eigenvalues and eigenvectors coalesce simultaneously \cite{Kato:1980}
leading to a defective eigenvalue problem.
Their existence is well documented in the context of non-Hermitian systems \cite{heiss:2012}, arising notably in the field of quantum physics \cite{moiseyev:2011, Cartarius:2009, Cartarius:2016, Egenlauf:2024}, optics and photonics \cite{miri:2019} with PT-symmetric systems \cite{bender:2018}.
Interesting contributions in the domain of mechanics can also be found
for the analysis of dynamical systems \cite{Triantafyllou:1991, Luongo:1995, Ghienne:2020, Even:2024} 
and their stability \cite{SeyranianBook:2003} and also in the context of 
guided acoustic waves \cite{Tester:1973, Shenderov:2000, Bi:2015, Bloch3D:2017, kanev2018, perrey-debain2022, Lawrie:2022, qiu2020}. %
In this latter case, it was demonstrated that EP assures maximum modal attenuation along the waveguide \cite{Bloch3D:2017, Perrey-Debain:2021}.


An eigenvalue problem perturbation refers to the study of how perturbations in a matrix or operator affect its eigenvalues and eigenvectors. This is a rich theme based on mathematical foundations \cite{Wilkinson:1988, Kato:1980}, with extensive applications in engineering where eigenvalue problems generally have a parametric dependence. Many examples can be found, the study of guided waves whereby wavenumbers, i.e. the eigenvalue of the problem, can be regarded as function of the frequency \cite{Krome:2016}, the characterization of the vibrational behaviour of structures presenting some uncertain parameters \cite{Adhikari:2007}, the computation of Campbell diagram of rotating machinery \cite{VeeringRotor} or even the prediction of flutter instability of airfoils \cite{SeyranianBook:2003} which depends on the rotation and the flow speed respectively.
When a single real parameter is slowly changed, some eigenvalues avoid to cross \cite{von1993verhalten}. This phenomenon called  \emph{veering} in structural dynamics \cite{Pierre:1988, Krome:2016, Ghienne:2020}, shows how eigenvalues interact and how perturbation may be sensitive when an EP exists in the complex-plane.

This variety of applications is well illustrated in the recent survey \cite{Mottershead:2020} reporting the state-of-the-art on the use of eigenvalues and eigenvectors derivatives required in the Taylor expansions.
If the analysis is generally limited to first and second order derivatives, recent works illustrate the interest for the computation of higher order derivatives. This can be done using the bordered matrix proposed in \cite{Andrew:1993} or the adjoint approach  \cite{Mensah:2020, Orchini:2021} using left and right eigenvectors, see for instance  \cite{Ghienne:2020, Nennig:2020, Mach:2023}.
In \cite{Ghienne:2020}, it was observed that the Taylor expansion becomes poorly convergent in situations where 
an exceptional point lies in the complex plane close to the real axis. %
In order to illustrate this, let us consider the simple case of a single complex-valued parameter $\nu$. In the vicinity of a double root characterized by the pair $(\lambda^*, \nu^*)$, the behavior of the two branches of solutions is given by a convergent Puiseux series \cite{Kato:1980} 
\begin{equation}\label{eq:Puiseux}
\lambda^\pm(\nu) = \lambda^* \pm d_1 (\nu-\nu^*)^{1/2} 
+ d_2 (\nu-\nu^*) \pm d_3 (\nu-\nu^*)^{3/2} + \ldots.
\end{equation}
where $d_i$ are known constants depending on the problem.
Consequently, the Taylor expansion of $\lambda^\pm$ calculated at an arbitrary value $\nu_0$ is expected to be limited to a certain radius of convergence which can not exceed $|\nu_0- \nu^*|$.
In order to circumvent the branch point singularity, two
auxiliary functions can be defined:
\begin{equation} \label{eq:gh}
g = \lambda^+ + \lambda^{-} \text{ and }
h= \left(\lambda^+ - \lambda^{-} \right)^2.
\end{equation}
The main advantage of these functions is their regularity in the neighborhood of $\nu^*$. Being regular, these functions can be expanded as convergent Taylor series. The reconstruction of the eigen pair $\lambda^\pm$ from the auxiliary functions, or through their Taylor expansion, can thus extend the applicability of Taylor expansions in the parameter space. This constitutes the main trick of this approach. The eigenvalues can be recovered as
\begin{equation}\label{eq:aaf}
\lambda^\pm= \frac{g\pm\sqrt{h}}{2} \approx \frac{\mathcal{T}_g \pm\sqrt{\mathcal{T}_h}}{2},
\end{equation}
where functions $\mathcal{T}_g$ and $\mathcal{T}_h$ are simply calculated from the original Taylor expansion of the pair eigenvalues $\lambda^\pm$ as described above. Not only, the reconstruction \eqref{eq:aaf}
is generally better than the Puiseux series, it also does not necessitate the \textit{a priori} knowledge of the double root location. The analyticity of $h$ has been exploited in many research works on EP \cite{Uzdin:2010, Hernandez:2005, Cartarius:2009}.
A general way to construct regular auxiliary functions is to see the link with the characteristic polynomial (although just the matrix trace here) for $g$ and its discriminant for $h$ which vanishes for multiple roots \cite{Cox:ideals}.
The idea is not totally new as this can be found in \cite{Fried:1989} or in \cite{Zheng:2022characteristic}.

A second advantage of the expression \eqref{eq:aaf} is to provide an efficient way to locate EPs in the complex plane. In this particular case, roots of $\mathcal{T}_h$ can be used to identify second-order EPs (call it $\mathrm{EP}_2$). 
In this regard, one can refer to a recent paper \cite{Nennig:2020} for a survey on different approaches proposed in the literature for the location of $\mathrm{EP}_2$
in the complex plane.
Some methods are based on a contour loop in the parameter space \cite{Hernandez:2005,Cartarius:2009,Uzdin:2010} and exploit the self-intersecting properties of the Riemann surface close to the EP.
The same idea combined to Gaussian-process-regression is used in \cite{Egenlauf:2024} to locate \EP{2} from specific eigenvalues of interest. %
A promising alternative is to exploit the analyticity of the matrix determinant as shown in \cite{Akinola:2014}.
Other approaches rely on the approximation of two eigenvalues using local physical model \cite{Cartarius:2009, Cartarius:2016} or through multipoint Padé approximants \cite{Lefebvre:2010,Landau:2015}.
The location of $\mathrm{EP}_2$ can also be found using metamodeling combining both response surface methodology and modal reduction  \cite{Gallina:2011} or by solving constrained minimization problems using Newton-type iterative methods \cite{Zsuzsanna:2018}.  
All these methods are local, use low order approximation, usually require an a priori physical information and may be difficult to apply when dealing with more than one parameter.

Global approaches which permit to obtain \textit{all} exceptional points also exist. They have been devised for the treatment of standard parametric eigenvalue problems which can also be recast into a sequence of multiple eigenvalue problems as shown in \cite{Jarlebring:2011,Muhivc:2014, Kiefer:2023}. The approach requires the treatment of matrices of size $M^2\times M^2$ ($M$ is the size of the original matrix system) making the method extremely prohibitive for large size matrices.

Exceptional points of order $n$ ($\mathrm{EP}_n$) require more than one parameter and their localization becomes more challenging than $\mathrm{EP}_2$ since the eigenvalue sensitivity increases with the number of parameters \cite{Kato:1980}.  From a computational point of view, their location in the parameter space is prone to rounding errors due to finite precision arithmetic \cite{Ryu:2021classification} and EPs often split into a cluster of nearly defective eigenvalues.
A possible route to obtain high order EP is to work on simple hierarchical system with explicit eigenvalue structure and to combine them \cite{Zhong:2020Hierarchical, wiersig2022revisiting}. Despite these difficulties, practical realization has been done \cite{Ding:2016emergence}.

To the best of the authors' knowledge, only a few generic methods have been developed for the location of high-order EPs for standard eigenvalue problems. Based on the versal theory, Mailybaev \cite{Mailybaev:2006} developed a local algorithm based on a Newton's method and a Schur decomposition at each step. 
Based on the same principle, Hernandez \cite{Hernandez:2017} proposes a more scalable algorithm requiring faster linear solve at each step.

Before concluding this introduction it could be mentioned that perturbation methods require the access to the operator derivatives.
When this knowledge is unavailable, though this is not the place for a full survey, black-box solvers also exist. On the one hand, efficient adaptive sampling strategy, based on continuation, has been proposed in \cite{Pradovera:2024}.
On the other hand, the original discrete problem can be projected onto a reduced subspace
whose basis is obtained from the solutions obtained by sampling the parametric space \cite{Sirkovic:2016, Ruymbeek:2019,Boffi:2024}.
Nonetheless, these model reduction approaches mostly focus on symmetric positive definite generalized eigenvalue problem and are often limited to the first eigenvalues.

The method proposed here relies on a partial characteristic polynomial and allows to build a (nearly) global approximation of the problem which is valid in large parametric domain.
The paper is organized as follows: the main theoretical ingredients of the method are explained in the next two sections. Section \ref{sec:theory} is dedicated to the construction of a partial characteristic polynomial based on a subset of eigenvalues. In Section \ref{sec:EPN}, the problem is transformed into a multivariate polynomial system for which numerical solutions can be sought using a panel of methods using either iterative, homotopy solvers or algebraic manipulations. In the last section three examples of increasing complexity are presented in order to show the numerical stability, the computational efficiency and complexity of the proposed method. In particular, specific features such as round-off errors arising from double-precision floating-point arithmetic and the computational burden caused by the calculation of high-order derivatives of eigenvalues are discussed.

\section{Partial characteristic polynomial}\label{sec:theory}

\subsection{Problem statement on an illustrative example}

\begin{figure}
  \begin{center}
    \includegraphics[width=0.4\textwidth]{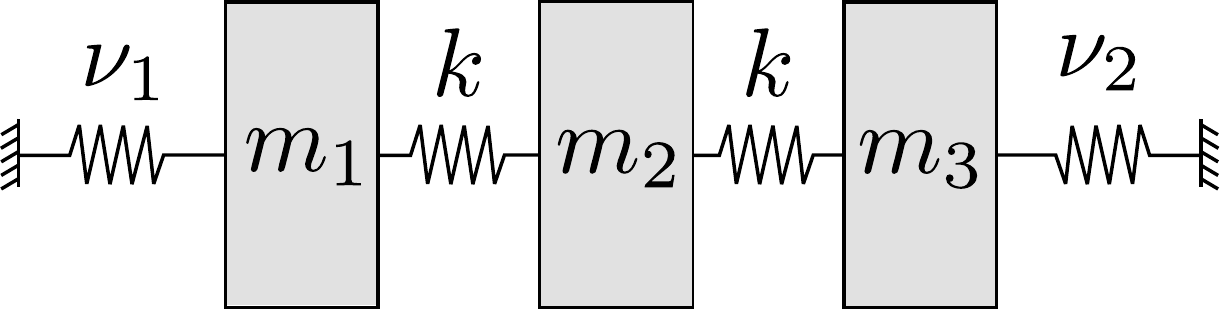}
    \caption{Toy model with two parameters.}\label{fig:toy}
  \end{center}
\end{figure}

We start with a toy model with $M=3$ degrees of freedom and 2 complex-valued parameters  $\nuv = (\nu_1, \nu_2)$  which represent the stiffness of the two springs at each end of the spring-mass system as shown in Fig.~\ref{fig:toy}. The imaginary part can be interpreted as viscoelastic contribution when working in the frequency domain. The matrix system takes the form of a generalized eigenvalue problem with
\begin{equation}
\Lv (\lambda, \nuv) =\mathbf{K}(\boldsymbol\nu) - \lambda\mathbf{M}.
\end{equation}
Here, we have $\lambda = \omega^2$ and the stiffness and mass matrix are given explicitly by
\begin{equation}
\mathbf{K}(\nuv) = \begin{pmatrix}k + \nu_1 & - k & 0\\
              - k & 2 k & - k\\
              0 & - k & k + \nu_2
              \end{pmatrix},
              \quad \mathrm{and} \quad 
              \mathbf{M} = \begin{pmatrix}m_1 &0 & 0\\
              0 & m_2  & 0\\
              0 & 0 & m_3
              \end{pmatrix}.
\end{equation}
To simplify the analysis, we take $k=m_1=m_2=m_3=1$, the associated characteristic polynomial has the general expression
\begin{equation}
\mathcal{P}(\lambda, \nuv) = 
(\lambda - \lambda_0)(\lambda - \lambda_1)(\lambda - \lambda_2)=
\lambda^3 + a_{2}(\nuv)\lambda^{2} + a_{1}(\nuv)\lambda + a_0(\nuv),
\end{equation}
with
\begin{subequations} \label{aaa}
\begin{align}
a_2(\nuv) &= -(\lambda_0 + \lambda_1 + \lambda_2 ) = - (\nu_1 + \nu_2 + 4),\\
a_1(\nuv) &=  \lambda_0 \lambda_1 \ +  \lambda_1 \lambda_2 + \lambda_2 \lambda_0  =
   \nu_1 \nu_2 + 3 \nu_1 + 3 \nu_2 + 3, \\
a_0(\nuv) &= -\lambda_0 \lambda_1 \lambda_2 = -  ( 2 \nu_1 \nu_2 + \nu_1 + \nu_2).
\end{align}
\end{subequations}
Because this is a cubic polynomial, roots have the explicit expression (the dependence on $\nuv$ has been omitted for clarity and a precise description of the cubic root is proposed by Baydoun \cite{Baydoun:2024}):
\begin{equation}
 \lambda_k =  -\frac{a_2}{3} + \j^k  \sqrt[3]{\frac{1}{2}
 \left(-q+\sqrt{\frac{\Delta}{27}} \right)}
 + \j^{-k}  \sqrt[3]{\frac{1}{2}
 \left(-q-\sqrt{\frac{\Delta}{27}} \right)},
 \end{equation}
where  $\j=\exp(\i 2\pi/3)$ and $\Delta$ is the discriminant of the polynomial given by
\begin{equation}
 \Delta = 4p^3 + 27 q^2 \quad \mathrm{with} \quad
 p=a_1 - \frac{a_2^2}{3} \quad \mathrm{and}  \quad q = \frac{2 a_2^3}{27} - \frac{a_1a_2}{3} +a_0.
 \end{equation}
At this point, two observations can be made: (i) expressions for the eigenvalues exhibit branch points singularities, one of them, when $\Delta=0$, is associated with the existence of a double (or triple) root. This means that the Taylor series calculated in the vicinity of a branch point is likely to have a limited domain of convergence ; (ii) %
polynomial coefficients $a_0$, $a_1$ and $a_2$ are expressed as a recombination of the eigenvalues. They are regular functions of $(\nu_1,\nu_2)$. This means that they can be well approximated, or exactly recovered is some situations such as the one illustrated here, by a multivariate power series. When dealing with large size matrices, only a small subset of eigenvalues is generally available. The question arises regarding the size of the convergence domain for coefficients associated with a partial characteristic polynomial containing a limited number of roots. The treatment of this difficult question would certainly require a theoretical analysis that goes beyond this paper. Nevertheless, one can shed some light using classical perturbation analysis in the vicinity of a repeated root \cite{Luongo:2015, Lancaster:2003}. Let us consider the situation of a double root $\lambda^*$ verifying 
\begin{equation} \label{doubleroot}
 \mathcal{P} (\lambda^*,\nuv^*) = \partial_\lambda \mathcal{P} (\lambda^*,\nuv^*) = 0.
 \end{equation}
We explore the neighborhood of the double root by selected curves $\nuv= \nuv(\varepsilon)$ with $\nuv(0)=\nuv^*$. It can be shown that\footnote{
The solution, given by a fractional power series, is found by expanding the characteristic equation about $\varepsilon=0$ (assuming  \eqref{doubleroot} and $\partial_\varepsilon \mathcal{P} (\lambda^*,\nuv^*) \neq 0$):
$$ \mathcal{P} (\lambda,\nuv(\varepsilon)) =
\varepsilon \partial_\varepsilon \mathcal{P}(\lambda^*,\nuv^*) 
+ \frac{(\lambda - \lambda^*)^2}{2}\partial^2_\lambda \mathcal{P}(\lambda^*,\nuv^*) +\ldots = 0.$$
where it is reminded that $\nuv(0)=\nuv^*$.} 
\begin{equation}
\lambda^{\pm} = \lambda^* \pm \varepsilon^{1/2}
\sqrt{-2 \left( \frac{\partial_\varepsilon \mathcal{P}}{\partial^2_\lambda \mathcal{P}} \right )_{(\lambda^*,\nuv^*)}} 
+ \mathcal{O} (\varepsilon) \quad (\mathrm{where} \, \, \partial_\varepsilon \mathcal{P} = 
\partial_{\nu_1} \mathcal{P} \partial_{\varepsilon} \nu_1 + \partial_{\nu_2} \mathcal{P} \partial_{\varepsilon} \nu_2).
 \end{equation}
In this case, the partial polynomial associated with  the two eigenvalues $\lambda^+$ and $\lambda^-$,
 \begin{equation}
(\lambda - \lambda^+)(\lambda - \lambda^-) =
\lambda^2 - (\lambda^+ + \lambda^-)\lambda  + \lambda^+ \lambda^-,
\end{equation}
has regular coefficients in a certain domain of the parameter space including branch points corresponding to the coalescence of the two selected eigenvalues, although each eigenvalue, taken separately, is singular at $\nuv^*$. Note that the regularity of the coefficients stems simply from the fact that $\lambda^+ +\lambda^- = g$ and $\lambda^+ \lambda^- = (g^2-h)/4$ where $g$ and $h$ are auxiliary functions introduced in \eqref{eq:gh}.
This trick has been exploited in \cite{Ghienne:2020} to deal efficiently with the variability of some parameters in structural dynamics. The work is based on the regularity of $g$ and $h$ and is limited to the coalescence of two eigenvalues with a single parameter. The purpose of the next section is to extend the idea to the treatment of an arbitrary number of parameters.

\subsection{Generalization to a subset of eigenvalues}
The starting point of this analysis is an important result given in \cite[Theorem 2.1]{Andrew:1993} stating that a  selected  eigenvalue $\lambda$ and its associated eigenvector are  necessarily analytic in a neighborhood of $\nuv_0=(\nu_{0,1}, \nu_{0,2}, \, \dots,\, \nu_{0,N})$:
\begin{equation}\label{eq:lda_Taylor}
\lambda(\nuv) = \lambda(\nuv_0) + \sum_{|\alphav| \geq 1} \frac{ \partial^{\alphav}\lambda(\nuv_0)}{\alphav!} (\nuv_0-\nuv)^{\alphav},
\end{equation}
as long as $\lambda(\nuv_0)$ is a simple eigenvalue\footnote{An eigenvalue is said \emph{semisimple} if its algebraic and geometric multiplicity are equal and \emph{simple} if its algebraic multiplicity is one.}. Here we use the multi-index notation $\alphav  = (\alpha_1, \dots, \alpha_N)$
with 
\begin{equation}\label{eq:alpha2}
\alphav !\equiv\alpha _{1}! \alpha _{2}!\cdots \alpha _{N}!,\quad 
\nuv^{\alphav}\equiv\nu_{1}^{\alpha _{1}}\nu_{2}^{\alpha _{2}}\cdots \nu_{N}^{\alpha _{N}}, \quad \mathrm{and}
\quad |\alphav |\equiv\alpha _{1}+\alpha _{2}+\cdots +\alpha _{N}.
\end{equation}
The notation for the derivatives is $$\partial^{\alphav} \lambda \equiv \partial^{\alpha_{1}}_{\nu_{1}} \partial^{\alpha _{2}}_{\nu_{2}}\cdots \partial^{\alpha _{N}}_{\nu_{N}} \lambda  =
\lambda^{(\alpha_1, \dots, \alpha_N)}.$$ 

Now, given $\Lambda$ a subset of $L$ selected eigenvalues $\lambda_l$ with $l = 0,\ldots,L-1$, we introduce the partial characteristic polynomial (PCP)  defined in a certain neighborhood of $\nuv_0$:
\begin{equation}\label{eq:Q}
\Q(\lambda, \nuv) = \prod_{l=0}^{L-1}  (\lambda - \lambda_l(\nuv)) = \sum_{k=0}^{L} a_k(\nuv) \lambda^k ,
\end{equation}
where coefficients are given explicitly by Vieta's formula
\begin{equation}\label{eq:ak}
a_k(\nuv) = (-1)^{k+1}\sum_{\mathbf{c} \in \mathcal{C}_k} \prod_{l \in \mathbf{c}} \lambda_l(\nuv),
\end{equation}
and the $L$-uplet $\mathbf{c}$'s are the elements of the combinatorial set:
$$\mathcal{C}_k = \lbrace \text{all ways to choose }(L  - k) \text{ distinct eigenvalues in } \Lambda\rbrace.$$The cardinality of this set is \cite[24.1.1]{Abramowitz:1965} $$\# \mathcal{C}_k=\binom{L}{L-k}.$$ 
As explained earlier, the recombination in \eqref{eq:Q} is expected to be regular in a larger domain of the parameter space than each eigenvalue taken separately. In practice, only truncated Taylor series are available and, for the sake of clarity and notational simplicity, we will introduce the variable $D$ as the highest order of derivation in each direction. The PCP coefficients are thus expanded as follows
\begin{equation}\label{eq:ak_Taylor}
{a_k}(\nuv) \approx \mathcal{T}_{a_k}(\nuv) = \sum_{\alpha_1=0}^D \cdots \sum_{\alpha_N=0}^D (\hat{a}_k)_{\alphav} \, (\nuv_0-\nuv)^{\alphav} \quad \mathrm{where} \quad (\hat{a}_k)_{\alphav} = \frac{\partial^{\alphav}a_k(\nuv_0)}{\alphav!}.
\end{equation}
Derivatives are calculated from
\begin{equation}
\partial^{\alphav} a_k = (-1)^{k+1}\sum_{\mathbf{c} \in C_k}\partial^{\alphav}  \left (\prod_{l \in \mathbf{c}} \lambda_l  \right ),
\end{equation}
where an explicit form can be obtained by application of the  multivariate Leibniz' rule (for the sake of clarity the formula is written in a general form by reordering the range of indices, here $\ell_{\mathrm{max}} = \# \mathbf{c}$ and $f_\ell$ is a function of $\nuv$):

\begin{equation}\label{eq:Leibniz_prod_mv}
 \partial^{\alphav}  \left(\prod _{\ell=1}^{\ell_{\mathrm{max}}}f_{\ell}\right)=
\sum _{|\kv^1 |=\alpha_1} \dots \sum _{|\kv^N |=\alpha_N} 
 {\binom{\alpha_1}{k^1_{1},\dots ,k^1_{\ell_{\mathrm{max}}}}} \dots {\binom{\alpha_N}{k^N_{1},\dots ,k^N_{\ell_{\mathrm{max}}}}}
\prod _{\ell=1}^{\ell_{\mathrm{max}}}
f_{\ell}^{(k^1_{\ell}, \dots, k^N_{\ell})},
\end{equation}
where $\kv^i$ ($i=1,\ldots,N$)  corresponds to the set of natural numbers $\kv^i= \lbrace k_1^i,\ldots, k_{\ell_{\mathrm{max}}}^i \rbrace$ and the summation accounts for all combinations verifying 
$$ |\kv^i | =k_1^i + k_2^i + \ldots +  k_{\ell_{\mathrm{max}}}^i = \alpha_i.$$
The multinomial coefficient in \eqref{eq:Leibniz_prod_mv} is given explicitly by 
$$ {\binom{\alpha_i}{k^i_{1},\dots ,k^i_{\ell_{\mathrm{max}}}}} = \frac{\alpha_i !}{\prod _{\ell=1}^{\ell_{\mathrm{max}}} k^i_\ell !}. $$
The domain of convergence of the PCP coefficients can be assessed via a simple estimator of the radius of convergence using the Cauchy-Hadamard's root test \cite[sec. 2.6]{Whittaker:1996} and by considering each parameter $\nu_i$ separately. This criterion states that for a power series (in terms of $\nu_i$), the radius of convergence is defined as 
\begin{equation}
\rho_{ki} = 1/\limsup_{q\rightarrow \infty}\abs{(\hat{a}_k)_{\betav_{i,q}}}^{1/q},
\end{equation}
where $$\betav_{i, q} =(0,\ldots,q,\ldots,0)$$
is a $N$-uplet with zeros everywhere except the component $i$ having the value $q$. This index basically selects the components of the Taylor series associated with parameter $\nu_i$.
Since only the first $D$ terms are available, the radius is estimated from least square fit on the truncated sequence. 
For convenience, we then use
\begin{equation}\label{eq:roottest}
    \rho_i = \min_{k} \rho_{ki},
\end{equation}
as a global estimator associated with parameter $\nu_i$.

\subsection{Computational aspects and complexity}\label{sec:algo_tricks}

\subsubsection{Computation of derivatives of eigenvalues (and eigenvectors)} 
In practice, derivatives of each selected eigenvalues can be recursively obtained by solving the system of linear equations with a bordered matrix~\cite{Andrew:1993} 
of size $(M+1)\times (M+1)$:
\begin{equation}\label{eq:Bordered}
\begin{bmatrix}
\Lv & (\partial_\lambda \Lv) \xv \\
\vv^t & 0
\end{bmatrix}
\begin{pmatrix}
\partial^{\alphav} \xv\\ \partial^{\alphav}\lambda
\end{pmatrix}
= \begin{pmatrix}
\Fv_{\alphav}\\ 0
\end{pmatrix},
\end{equation}
where the right-hand side (RHS) vector $\Fv_{\alphav}$ contains terms arising from previous order derivatives and all necessary details are briefly reminded in~\ref{app:derivative}.
The approach is very stable numerically and it is noteworthy that the computation time for the construction of the RHS may become long because of combinatoric operations especially when the number of parameters increases (this is discussed in Sec.~\ref{sec:time}).
Depending on the matrix size and storage, \eqref{eq:Bordered} can be run in parallel for each eigenvalue in $\Lambda$.
 The algorithm is implemented in \texttt{EasterEig} (v2) \cite{EasterEig}, an open source framework dedicated to perturbation of eigenvalue problems.
The python implementations allows to work with full, sparse and parallel sparse matrix using numpy \cite{numpy}, scipy.sparse \cite{scipy} or PETSc/PETSc4py (v3.22) \cite{Petsc, Mumps, Dalcin:2011} and SLEPc/SLEPc4py \cite{SLEPc, Dalcin:2011}. Note that if the technique breaks down in the specific case where $\lambda(\nuv_0)$ is a semisimple eigenvalue, alternative approaches have been developed \cite{Lancaster:2003, VanDerAa:2007, Orchini:2021, Qian:2017}.

\subsubsection{Computation of the truncated Taylor expansion of the PCP}\label{sec:hireachical}
In order to limit the number of terms in Eq.~\eqref{eq:Leibniz_prod_mv}, it is better to compute Eq.~\eqref{eq:Leibniz_prod_mv} hierarchically. For instance if we consider the product of $\ell_{\mathrm{max}}=8$ functions $f_1 \dots f_8$, it can be regarded either as the product of four functions $g_1 g_2g_3g_4$ or the product of two functions $h_1 h_2$ at top level, as illustrated in
\begin{equation}\label{eq:recursive_Leibniz_prod}
 \partial^{\alphav}  \left(\prod _{\ell=1}^{8}f_{\ell}\right) =   \partial^{\alphav}  \bigl(\underbrace{\underbrace{f_1 f_2}_{g_1} \underbrace{f_3 f_4}_{g_2}}_{h_1} \underbrace{\underbrace{f_5 f_6}_{g_3} \underbrace{f_7 f_8}_{g_4}}_{h_2} \big) =  \partial^{\alphav}  \left(\prod _{\ell=1}^{2}h_{\ell}\right).
\end{equation}
Here, $f_i$ corresponds to the previously computed partial derivative of each eigenvalue obtained with Eq.~\eqref{eq:Bordered} for instance.
Note that $\ell_{\mathrm{max}}$ can be an  odd number by considering the grouping of 3 functions  at a lower level. 
The proposed implementation uses a queue where functions involved at the lower-order level are treated as pairs or triplets and then the derivative for the pair (or triplet) is enqueued. The process is repeated until the queue contains only one element. The speed up with this hierarchical approach can be very  substantial since the computation time grows linearly with the number of eigenvalues $L$ whereas direct computation of Eq.~\eqref{eq:Leibniz_prod_mv} scales as $\mathcal{O}(L!)$. For instance, if 18 eigenvalues are retained, the hierarchical approach achieves a speed-up of approximately 10,000 times.
Clearly, without this acceleration process the method would be intractable from a computational point of view. Additionally, it  also limits the accumulation of round-off errors in the summation.
The main algorithmic steps for the construction of the PCP are summarized in Algorithm~\ref{alg:PCP}.

\subsubsection{Recovery of eigenvalues from the PCP}  

Given an arbitrary parameter vector $\nuv$ chosen in a certain neighborhood of the evaluation point $\nuv_0$, the first step is to evaluate $\mathcal{T}_{a_k}(\nuv)$  using the multivariate Hörner scheme. This yields an $L$-order univariate polynomial whose roots obeying
\begin{equation} 
\sum_{k=0}^{L} \mathcal{T}_{a_k}(\nuv)   \lambda^{k}(\nuv) = 0,
\end{equation}
are expected to be good approximations of the $L$ eigenvalues belonging to the selected subset.
If closed-form expressions are available for low order polynomials, the companion matrix method is used for the other cases. The complexity of this algorithm grows as $\mathcal{O}(L^3)$ \cite[Chap. 7]{Golub:2013} and works generally well for degrees lower than 20.
Note that in practice, the number of selected eigenvalues $L$ is much smaller than the matrix size $M$ and the computational cost for eigenvalue recovery is negligible when compared to a direct computation of the eigenvalue from the original algebraic system \eqref{eq:PVP}.
We note in passing that if more eigenvalues
have to be tracked, several subsets $\Lambda$ (possibility with partial overlap) can be considered.

\begin{algorithm}[htb]
\algrenewcommand\algorithmicrequire{\textbf{Input:}}
\algrenewcommand\algorithmicensure{\textbf{Output:}}
\algrenewcommand{\algorithmiccomment}[1]{\hfill{\color{black!60}$\triangleright${#1}}}
\begin{algorithmic}[1]
\Require Given a discrete operator $\Lv$ for which all partial derivatives with respect to $\nuv$ are available and $\nuv_0$.
\Ensure The partial characteristic polynomial expressed in the form \eqref{eq:Q}.
\State Solve the eigenvalue problem \eqref{eq:PVP} for a certain number of eigenvalues for $\nuv=\nuv_0$.
\State Create $\Lambda$, a subset of $L$ selected eigenvalues $\lambda_l$ with $l = 0,\ldots,L-1$.
\For{each $\lambda_l \in \Lambda$} \Comment{May be run in parallel}
	\State Solve the linear system \eqref{eq:Bordered} for each RHS $\Fv_{\alphav}$ 
\EndFor	
\item Compute the PCP coefficient $a_k(\nuv)$ by applying the recursive version of Eqs.~\eqref{eq:ak} to \eqref{eq:Leibniz_prod_mv}.
\State Estimate the radii of convergence $\rho_i$, $i=1,\,\dots,\, N$ with Eq.~\eqref{eq:roottest}
\end{algorithmic}
\caption{Construction of the partial characteristic polynomial.}\label{alg:PCP}
\end{algorithm}

\section{Location of high-order exceptional points}\label{sec:EPN}
\subsection{Exceptional point algebraic problem}
For non Hermitian problems, exceptional points (EP) correspond generically to situations where the characteristic polynomial
\begin{equation}
    \mathcal{P}(\lambda, \nuv) = \det \left (  \Lv (\lambda, \nuv) \right )
\end{equation}
has repeated roots. Given $N$ independent complex-valued parameters $\nu_i$ ($i=1,\ldots,N$), we are interested in high-order EPs satisfying
\begin{equation}
    \mathcal{P}(\sv^*) = \partial_\lambda\mathcal{P}(\sv^*)
    = \partial_\lambda^{N-1}\mathcal{P}(\sv^*)= 0,
\end{equation}
where for convenience, we called $\sv=(\lambda, \nuv)$ the extended parameter vector and symbol (*) refers to EP.
The explicit expression of $\mathcal{P}$ is unavoidable but one can still use the truncated version of the partial polynomial $\mathcal{Q}$. The system of equations becomes 
\begin{equation} 
 \partial^i_\lambda\mathcal{Q}(\sv^*) \approx \sum_{k=i}^{L} \mathcal{T}_{a_k}(\nuv^*) \binom{k}{i}  (\lambda^*)^{k-i} = 0,
\quad i=0,\ldots,N-1.
\end{equation}
This defines a typical multivariate polynomial system which can be recast in the compact form
\begin{equation} \label{eq:EP_system}
\Sv(\sv^*) = \mathbf{0}
\end{equation}
and has been the topic of intense research both from a mathematical and computational point of view. At this point, few remarks can be made: first, in most situations, the system 
has discrete solutions whose number can be obtained from the Bezout Number \cite{Wise:2000}. Second, not all these solutions are EPs. Spurious roots can appear due to the truncation of the Taylor series or due to the existence of semisimple eigenvalues which are generally observed when dealing with symmetric configurations. What distinguishes EPs from semisimple eigenvalues can be identified by using perturbation
analysis \cite{Luongo:2015, Welters:2011}.

\subsection{Multivariate polynomial system resolution}
The multivariate polynomial system can be solved using different techniques which can be listed below:

\begin{itemize}
\item \emph{Iterative solvers}. These methods are based on the use of the Newton-Raphson (NR) algorithm which needs  initial guesses  $\sv_{\mathrm{i.\,g.}}$ and thus requires the definition of a grid of points $\mathcal{G}$ in the parameter space to ensure that most roots can be found.
The grid $\mathcal{G}$ in the extended parameter space contains the combination of all eigenvalues of $\Lambda$ and the parameters
\begin{equation}\label{eq:grid}
    \nuv_{\mathrm{i.\,g.}} = \nuv_0 + (p_1 +\i q_1,\, p_2 +\i q_2,\dots, p_N +\i q_N),
\end{equation}
where $p_i$ (resp. $q_i$) range from $-\rho_i/2$ to $+\rho_i/2$ ($i=1,\, \dots,\, N$) with 3 or 4 equispaced values.

The NR algorithm is usually fast but tends to diverge for some initial guesses. 
The Levenberg-Marquard (LM) algorithm originally proposed for non-linear least square problems offers better performances. It contains a damping factor ensuring convergence even if the initial guess is far from the solution (see MINPACK \cite{minpack}/scipy.optimize.root \cite{scipy}). The LM method, which requires to split the complex-valued parameters into real and imaginary parts, will be used in most numerical examples shown in this work.

\item \emph{Homotopy solvers} \cite{Wise:2000, pypolsys}.
These methods allow to find all the solutions of Eq.~\eqref{eq:EP_system} numerically. The basic idea is to exploit analyticity properties in order to move gradually towards the true solutions from the solutions of a trivial and similar problem. Because of the existence of many spurious solutions, homotopy solvers should be applied to small problems with a limited number of roots. 
Such solver provides a fast estimate of Bézout's number before starting to solve the actual system. Here we use \cite{Wise:2000, pypolsys}.

\item Another approach is to work on the resultant \cite[p. 161]{Cox:ideals} which is defined as the determinant of the Sylvester matrix allowing to check if two polynomials share the same roots. In the univariate case, the Sylvester matrix is strongly related to the discriminant which also vanishes for multiple roots. It has been successfully used in \cite{Nennig:2020, Ghienne:2020} but here results were found to be less accurate for the test cases considered in this paper. For the sake of completeness, the main ingredients of the method are recalled in~\ref{app:discriminant}). 

\end{itemize}

\subsection{Spurious roots filtering}
Spurious roots are likely to be very sensitive to round-off errors and to the truncation of the Taylor series. Thus, spurious roots could be filtered out by comparing two sets of solutions calculated at two successive truncation orders as in \cite{Nennig:2020} but this may prove  costly when dealing with many parameters. Instead, we propose to introduce a sensitivity indicator which corresponds to a single NR correction step starting from the solution $\sv$ obtained using the maximum order of derivation $D$ in the Taylor expansion \eqref{eq:ak}:
\begin{equation}\label{eq:sensitivity}
    \delta = \Vert\mathbf{J}^{-1} \Sv(\sv)\Vert_2,
\end{equation}
where both the multivariate system $\Sv$ and the associated Jacobian matrix $\mathbf{J}=\frac{\partial \Sv}{\partial \sv}$ are computed using $D-1$ derivatives in the expansion. This approach is closely linked to Wilkinson's definition of polynomial roots condition number \cite[(4.2)]{Wilkinson:1988}.
The main algorithmic steps to locate EP are summarized in Algorithm~\ref{alg:EPN}.

\begin{algorithm}[htb]
\algrenewcommand\algorithmicrequire{\textbf{Input:}}
\algrenewcommand\algorithmicensure{\textbf{Output:}}
\algrenewcommand{\algorithmiccomment}[1]{\hfill{\color{black!60}$\triangleright${#1}}}
\begin{algorithmic}[1]
\Require A partial characteristic polynomial expressed in the form  \eqref{eq:ak_Taylor} at $\nuv_0$.
\Ensure The filtered discrete set $\mathcal{S}$ of solutions $\sv^*$ corresponding to $\mathrm{EP}_N$.
\State Estimate the radii of convergence $\rho_i$, $i=1,\,\dots,\, N$ with \eqref{eq:roottest}
\State From the system \eqref{eq:EP_system}, estimate the Bézout number $B$ \Comment{Use \cite{pypolsys}}
\If{$B > B^{\mathrm{max}}$}
\State Define the grid $\mathcal{G}$ in the extended parameter space using Eq.~\eqref{eq:grid}%
\Comment{Based on $\rho_i$, $i=1,\,\dots,\, N$}
\For{each point $\sv_{\mathrm{i.\, g.}} \in \mathcal{G}$}
	\State Run LM (or NR) solver starting from $\sv_{\mathrm{i.\, g.}}$
	\State Append the solution to the solution set $\mathcal{S}$
\EndFor
\Else
	\State Run homotopy solver to find the solution set $\mathcal{S}$
\EndIf
\State Filter spurious roots from $\mathcal{S}$ using Eq.~\eqref{eq:sensitivity}
\end{algorithmic}
\caption{Location of exceptional points of order $N$ (\EP{N}).}\label{alg:EPN}
\end{algorithm}

\section{Examples}\label{sec:examples}

The three following examples of growing complexity are chosen in order to illustrate and analyze several specific features of the method such as accuracy and robustness. The first one corresponds to the very simple model of Section \ref{sec:theory}. The second example deals with guided waves in a bi-dimensional acoustic duct with lined walls. The last example gives an application of the method in room acoustics.

\subsection{A toy model with 3 degrees of freedom}\label{sec:toy-model}%

We consider the toy model introduced earlier in section \ref{sec:theory}. In this scenario, the application of the bordered matrix is simplified since only first derivatives of the operator are non-zero and
\begin{align}
&\partial_{\nu_1} \Lv= \begin{pmatrix}1 & 0 & 0\\
              0 & 0 & 0\\
              0 & 0 & 0
              \end{pmatrix}, 
&\partial_{\nu_2} \Lv &= \begin{pmatrix}0 & 0 & 0\\
              0 & 0 & 0\\
              0 & 0 & 1
              \end{pmatrix},
&\partial_\lambda \Lv=-\Mv.
\end{align}

The fact that the matrix size is very small allows to consider the 3 eigenvalues of the problem and the complete (and not partial) characteristic  polynomial can be recovered with the Algorithm~\ref{alg:PCP}. By choosing $\nuv_0=(1,\, 1)$  and by taking $D=7$ as the maximum order of derivation with respect to the parameters, we find
\begin{subequations} \label{a_expression}
\begin{align}
 \mathcal{T}_{a_0}(\nuv)  &= -1.0 \nu_1 - 1.0 \nu_2 - 4.0,  \\
 \mathcal{T}_{a_1}(\nuv)  &=   1.0 \nu_1 \nu_2 + 3.0 \nu_1 + 3.0 \nu_2 + 3.0, \\
 \mathcal{T}_{a_2}(\nuv)  &= -2.0 \nu_1 \nu_2 - 1.0 \nu_1 - 1.0 \nu_2 +0.0,
\end{align}
\end{subequations}
which is in perfect agreement with the exact expressions in \eqref{aaa}. 
Note that (i) all coefficients, including those which are not displayed here as they should be equal to zero, are correct up to $10^{-13}$ which is close to machine precision and (ii) 
similar results are obtained by choosing another evaluation point $\nuv_0=(100,\, 50+50 \i)$. The fact that the solution is independent of the evaluation point $\nuv_0$ stems from the fact that \textit{all} eigenvalues of the problem have been considered in the recovery process and the dependency with respect to the parameters is quite simple, here linear. 

\begin{figure}
  \begin{center}
    \subfigure[$\log_{10} |(\hat{\lambda}_{0})_{\alphav}|$]{\includegraphics[width=0.3\textwidth]{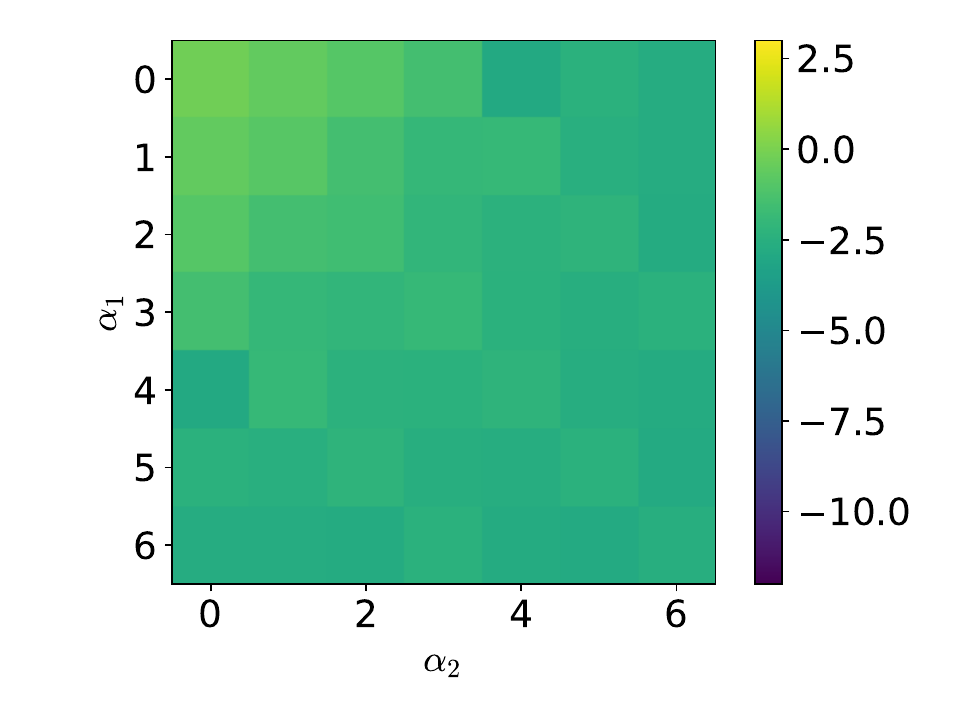}}
    \subfigure[$\log_{10} |(\hat{\lambda}_{1})_{\alphav}|$]{\includegraphics[width=0.3\textwidth]{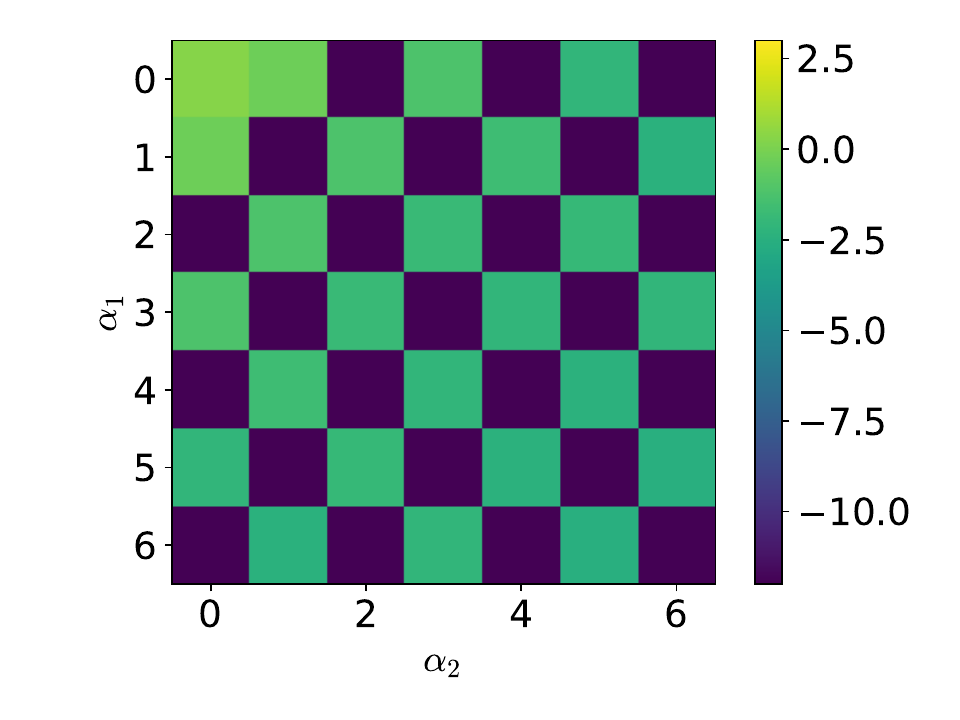}}
    \subfigure[$\log_{10} |(\hat{\lambda}_{2})_{\alphav}|$]{\includegraphics[width=0.3\textwidth]{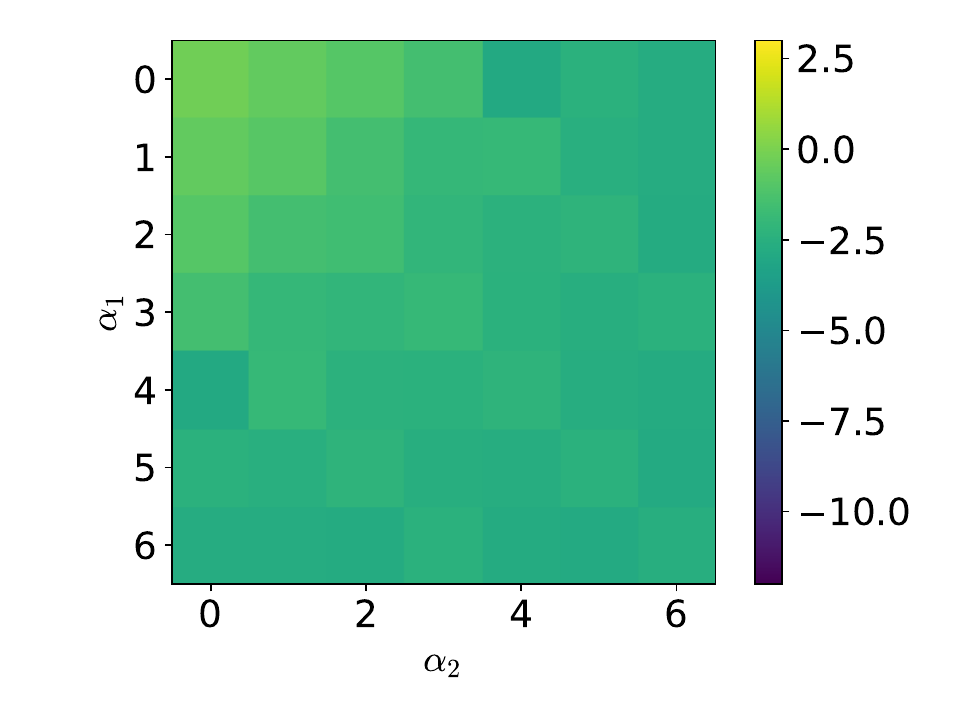}}
   \caption{Taylor coefficients of the eigenvalue $(\hat{\lambda}_k)_{\alphav} = \partial^{\alphav}\lambda_k(\nuv_0) / \alphav!$ with $k=0,\, 1,\, 2$. Computed with $\nuv_0=(1, 1)$ and $D=7$.}
      \label{fig:coef_lda}
  \end{center}
\end{figure}
\begin{figure}
  \begin{center}
    \subfigure[$\log_{10} |(\hat{a}_{2})_{\alphav}|$]{\includegraphics[width=0.3\textwidth]{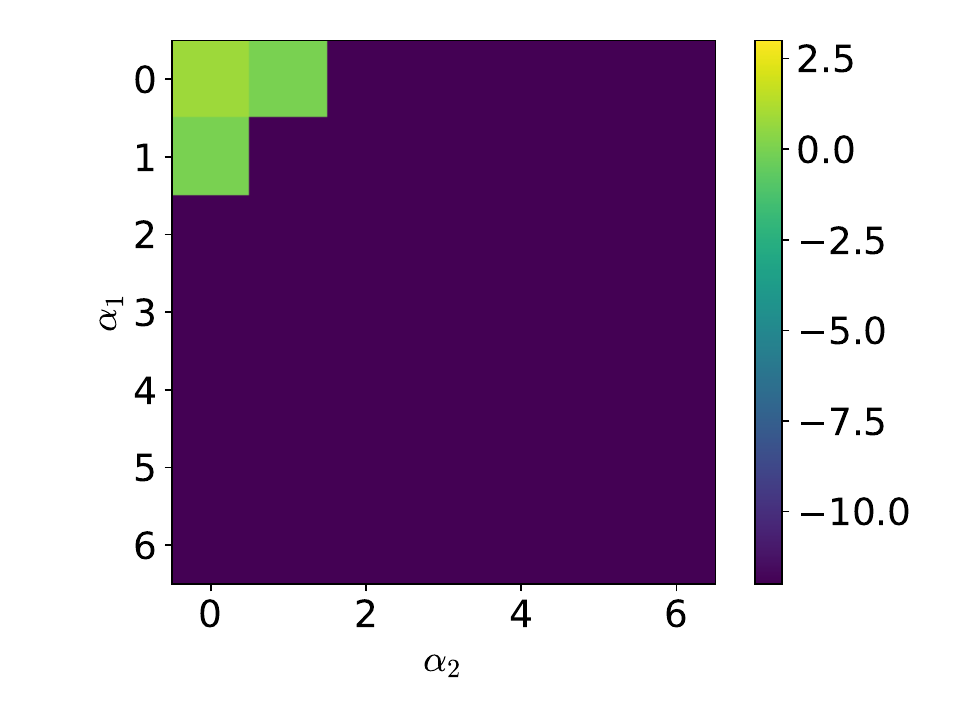}}
    \subfigure[$\log_{10} |(\hat{a}_{1})_{\alphav}|$]{\includegraphics[width=0.3\textwidth]{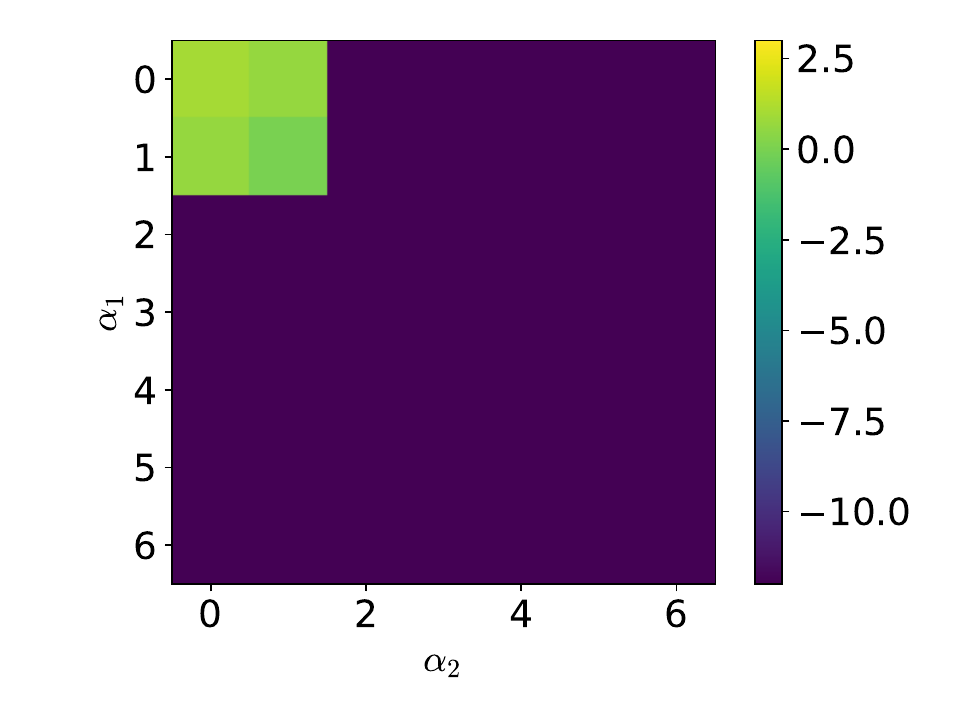}}
    \subfigure[$\log_{10} |(\hat{a}_{0})_{\alphav}|$]{\includegraphics[width=0.3\textwidth]{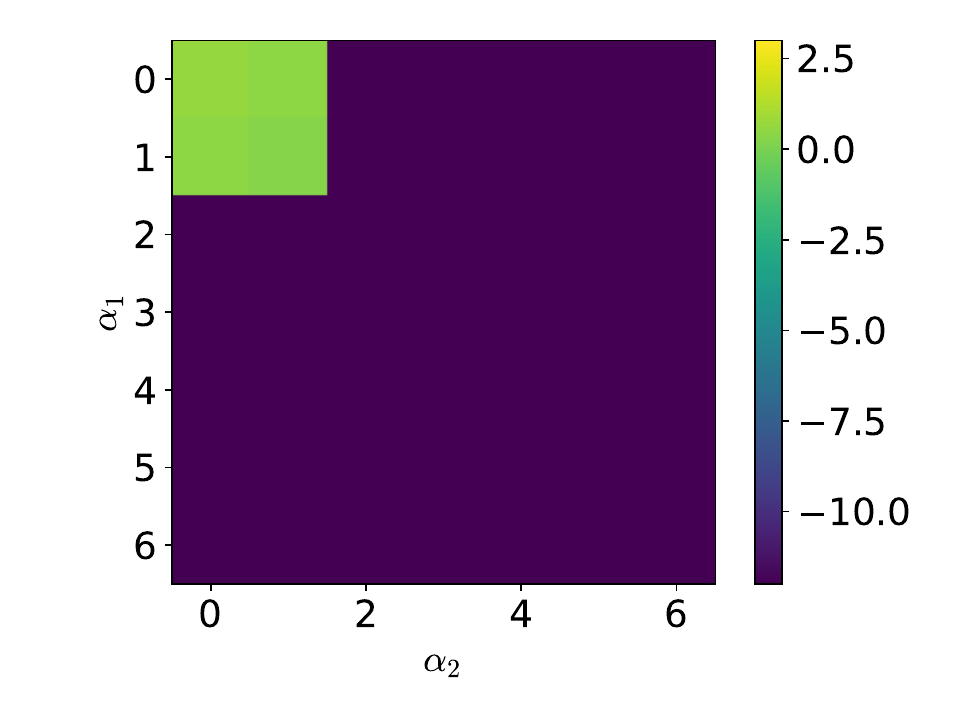}}
   \caption{Taylor coefficients of the PCP coefficients defined in \eqref{eq:ak_Taylor} computed with $\nuv_0=(1, 1)$ and $D=7$.}
   \label{fig:coef_a}
  \end{center}
\end{figure}
At this point, it is instructive to compare the amplitudes of the Taylor coefficients for the 3 eigenvalues, illustrated in Fig.~\ref{fig:coef_lda}, to those of the PCP coefficients, i.e. $|(\hat{a}_k)_{(\alpha_1,\alpha_2)}|$ for $k=0,1,2$ and $\alpha_i = 0,\ldots,D$ ($i=1,2$), illustrated in Fig.~\ref{fig:coef_a}. %
It can be seen that the Taylor coefficients of all eigenvalues seem to diverge whereas PCP coefficients are well-behaved which is in line with the exact expression \eqref{a_expression}. In theory this means that the radius of convergence, as presented earlier, should
be infinite. This is confirmed by Eq.~\eqref{eq:roottest} which shows that $\rho_i\approx 1000$ ($i=1,\, 2)$.

In order to locate the third order exceptional points $\mathrm{EP}_3$, we need to solve the following multivariate polynomial system (the star symbol is omitted here for clarity)
\begin{equation}\label{eq:EP_system_3dof}
\left\lbrace
\begin{aligned}
- 2 \lambda - \nu_1 - \nu_2 + \left(\lambda + 2\right) \left(\lambda + \nu_1 + 1\right) \left(\lambda + \nu_2 + 1\right) - 2 &= 0, \\
\left(\lambda + 2\right) \left(\lambda + \nu_1 + 1\right) + \left(\lambda + 2\right) \left(\lambda + \nu_2 + 1\right) + \left(\lambda + \nu_1 + 1\right) \left(\lambda + \nu_2 + 1\right) - 2 &= 0, \\
6 \lambda + 2 \nu_1 + 2 \nu_2 + 8 &= 0.
\end{aligned}
\right.
\end{equation}
This system has the remarkable property that it can be solved analytically using the Gröebner basis \cite[Chap. 2]{Cox:ideals} (this can
be seen as a generalization of the Gaussian elimination for system of polynomials). %
The Gröebner basis, computed over the integer field yields,
\begin{equation}
\left\lbrace
\begin{aligned}
63\lambda + 2\nu_2^5 - 10\nu_2^4 + 43\nu_2^3 - 89\nu_2^2 + 117\nu_2 -  189 &= 0, \\
21 \nu_1 + 2 \nu_2^5 - 10 \nu_2^4 + 43 \nu_2^3 - 89 \nu_2^2 + 138 \nu_2 - 105 &= 0, \\
(\nu_2^2 - 3 \nu_2 + 9)(\nu_2^2 - 2 \nu_2 + 3) (\nu_2^2 - \nu_2 + 7) &= 0 .
\end{aligned}
\right.
\end{equation}
Finally, 3 conjugate pairs of $\mathrm{EP}_3$ are found :

\begin{subequations}
\begin{align}
\sv^*_1 &= \left(2, 1 \mp \sqrt{2}\i, 1 \pm \sqrt{2}\i\right), \\
\sv^*_2 &=  \left(2 \pm \sqrt{3}\i, \frac{1 \pm 3\sqrt{3}\i}{2}, \frac{3\pm 3\sqrt{3}\i}{2} \right), \\
\sv^*_3 &= \left(2 \pm \sqrt{3}\i, \frac{3\pm 3\sqrt{3}\i}{2}, \frac{1\pm 3\sqrt{3}\i}{2} \right).
\end{align}
\end{subequations}

Here, the repeated values of the  solutions stems simply from the fact that the configuration is symmetric and the two parameters $\nu_1$ and $\nu_2$ play a similar role:
 $ \mathcal{P}(\lambda, \nu_1, \nu_2) =  \mathcal{P}(\lambda, \nu_2, \nu_1)$. It can be noticed that the first eigenvalue, $\lambda=2$, is real-valued due to the fact that the two parameters are complex conjugate.
The homotopy solver is used in Algorithm~\ref{alg:EPN} and the computed solutions are given in Tab.~\ref{tab:3dof_roots}. Here again, the agreement is excellent with an (absolute) error that does not exceed $10^{-8}$ and the sensitivity indicator $\delta$ is below $10^{-11}$. All the other spurious roots are easily discarded as the related sensitivity indicator is significantly bigger than 1 (up to $10^9$) and this shows that the filtering strategy is reliable. Before we leave this section, it is instructive to give a physical description of the nature of the solution associated with an $\mathrm{EP}_3$. A solution of interest is obtained from $\omega^* = \sqrt{2+\sqrt{3}\i} \approx 1.5241+0.5682 \i$ and
\begin{equation}
  \qv(t)= \mathrm{Re} \left ( \xv^* \e^{(\i 1.5241 - 0.5682)t }\right ),
\end{equation}
where $\xv^*$ is the unique eigenvector of the eigenvalue problem and $\qv$ is the vector of generalized coordinates which is exponentially decaying due to the energy dissipation stemming from the imaginary part of the spring constants. It can be shown that the decay rate is optimal (at least in a certain neighborhood of $\nuv^*$) for the specific choice $\nuv^*=(\nu_1^*,\nu_2^*)$. A practical  realization of this phenomenon for the classical coupled pendulum can be found in a recent paper \cite{Even:2024}.

\begin{table}
\resizebox{\columnwidth}{!}{%
\begin{tabular}{cccc}
\hline 
$\lambda^*$ & $\nu_1^*$ & $\nu_2^*$ & $\delta$\\
\hline
2.0 + 7.627560048174e-15$\i$ & 0.99999999999999 + 1.4142135623731$\i$ & 1.0 -1.4142135623731$\i$ & 2.63e-14 \\
2.0 -1.1580887031809e-13$\i$ & 0.99999999999999 -1.4142135623733$\i$ & 1.0 + 1.414213562373$\i$ & 2.42e-13 \\
2.0 -1.7320508075689$\i$ & 1.5 -2.5980762113536$\i$ & 0.5000000000001 -2.5980762113531$\i$ & 5.29e-13 \\
2.0 -1.7320508075689$\i$ & 0.50000000000017 -2.5980762113531$\i$ & 1.4999999999999 -2.5980762113537$\i$ & 5.83e-13 \\
2.0 + 1.7320508075689$\i$ & 0.50000000000013 + 2.5980762113531$\i$ & 1.4999999999999 + 2.5980762113537$\i$ & 6.04e-13 \\
1.9999999999991 + 1.7320508075673$\i$ & 1.4999999999983 + 2.598076211353$\i$ & 0.499999999998 + 2.5980762113503$\i$ & 4.47e-12 \\
\hline
\end{tabular} 
}
\caption{Solution obtained using Algorithm~\ref{alg:EPN} with homotopy solver starting from $\nuv_0=(1, 1)$ with $L=3$. To limit the number of solutions, the PCP is truncated to $D=5$. The Bézout's number is 192.}
\label{tab:3dof_roots}
\end{table}

\subsection{A bi-dimensional acoustic waveguide with arbitrary impedance boundary conditions}\label{sec:guide}

\begin{figure}
    \centering
    \includegraphics[width=0.5\textwidth]{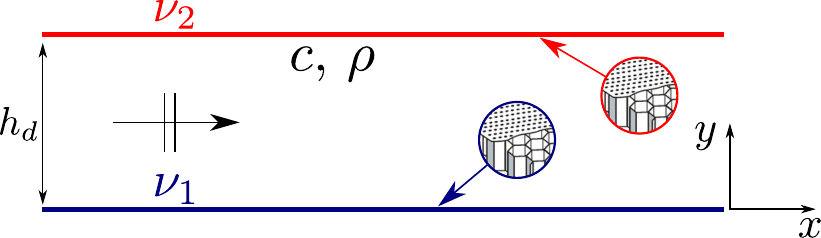}
    \caption{Acoustical waveguide lined with 2 admittance boundary condition.}
    \label{fig:duct}
\end{figure}

We consider the bi-dimensional acoustic waveguide of infinite length with a 1D cross-section of unit height, i.e. $h_d=1$, depicted in Fig.~\ref{fig:duct}.
The acoustic pressure satisfies the Helmholtz equation (we take the time convention $\e^{-\i \omega t}$ where $\omega$ is the angular frequency and we  put $c=1$ where $c$ is the sound speed)
\begin{equation}\label{eq:Helmholtz}
\Delta p + \kappa^2 p =0,
\end{equation}
with the free field wavenumber $\kappa=\omega/c$.
On the duct walls, a local boundary condition  is prescribed:
\begin{equation} \label{eq:BC}
\partial_y p = - {\nu_1} p, \text{ at } y=0 \quad \text{and,} \quad \partial_y p =  {\nu_2} p, \text{ at } y=1,
\end{equation}
where the normalized wall admittance $\nu_1$ and $\nu_2$ are the two complex-valued parameters of the problem which model acoustics treatments. Following classical modal analysis, duct acoustic modes are sought in the separable form 
\begin{equation}\label{eq:Zwg}
p(x,y) = \phi(y) \e^{\i (\beta x - \omega t)}.
\end{equation}

This example has been the subject of a complete analytical study in \cite{perrey-debain2022} showing the existence of second and third order exceptional points $\mathrm{EP}_2$ and $\mathrm{EP}_3$ whose exact location in the complex plane can be found and serve as a reference solution.
A FEM discretization (using 200 linear elements) of the weak formulation 
\begin{equation*}
- \int_0^1 \partial_y \psi  \partial_y \phi \ud y + (\kappa^2 - \beta^2)\int_0^1 \psi \phi \ud y 
+ \nu_1 \psi(0)  \phi(0) +  \nu_2 \psi(1)  \phi(1)=0,
\end{equation*}
yields the  parametric generalized eigenvalue  problem, (here $\lambda=\beta^2$  and ${\nuv}=({\nu_1}, {\nu_2})$):
\begin{equation}\label{eq:PVPz}
\Lv \big(\lambda({\nuv}), {\nuv} \big) \xv({\nuv})  =  \big(-\Kv + (\kappa^2-\lambda({\nuv})) \Mv  +{\nu_1} \boldsymbol{\Gamma}_1 + {\nu_2} \boldsymbol{\Gamma}_2 \big)  \xv({\nuv}) =\mathbf{0},
\end{equation}
where $\Kv$ and $\Mv$ are the standard stiffness and mass matrix respectively.

In this example, the number of eigenvalues is too large and a partial polynomial must be considered. In this context, we are interested in the effect of the number of eigenvalues retained in the subset $\Lambda$ on the quality of the eigenvalue  reconstruction. We consider the trajectory in the parameter space 
\begin{equation}\ 
\nuv(\epsilon) = \nuv_0 + \epsilon\ \e^{0.3\i} \begin{pmatrix}
    1\\
    1
\end{pmatrix}, \quad \mathrm{where} \quad \epsilon\in[0, 10] \quad \mathrm{and} \quad 
\nuv_0=(4.76715+7.01265\i, 2.470   +2.89872\i)
\end{equation}
and define the global eigenvalue reconstruction error
\begin{equation}\label{eq:error_wg}
E(\epsilon) = \max_{l=0,\ldots,L-1} \abs{\lambda_l^\text{dir.}(\nuv(\epsilon))  - \lambda_l^\text{pcp}(\nuv(\epsilon))} \, \text{for all $\lambda_l \in \Lambda$}, 
\end{equation}
where $\lambda_l^\text{dir.}$ are computed directly with the eigensolver and 
$\lambda_l^\text{pcp}$ are reconstructed from the PCP obtained at $\nuv_0$. 
The eigenvalues $\lambda_l^\text{dir.}$ and $\lambda_l^\text{pcp}$ are paired by solving a \emph{linear sum assignment} problem to find the best global match between both eigenvalue sets.

\begin{figure}
  \begin{center}
    \subfigure[]{\includegraphics[width=0.48\textwidth]{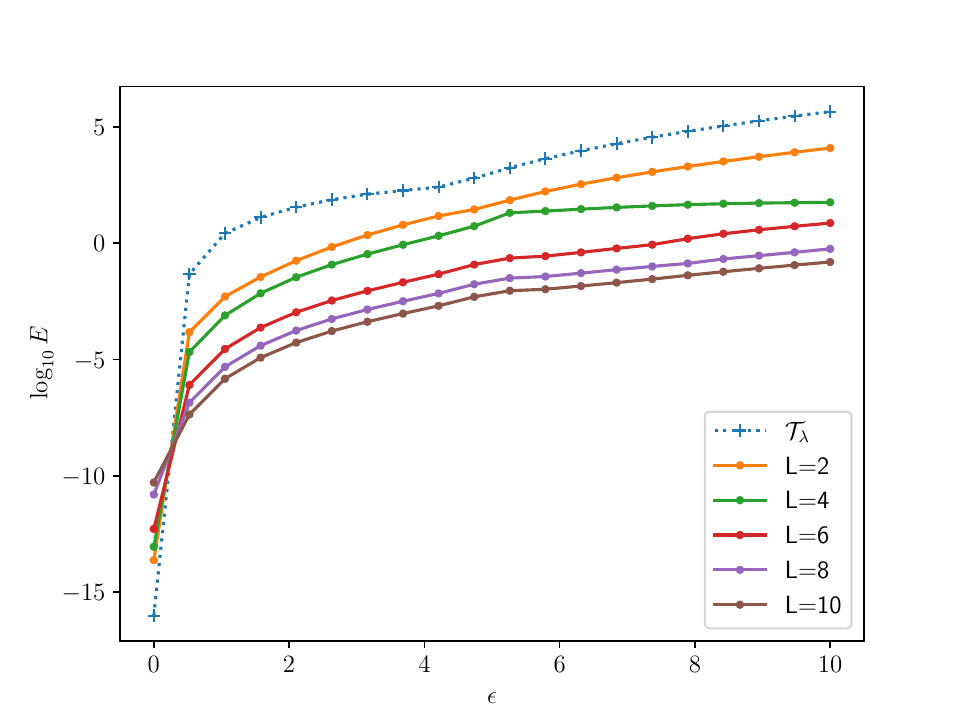}}
    \subfigure[]{\includegraphics[width=0.48\textwidth]{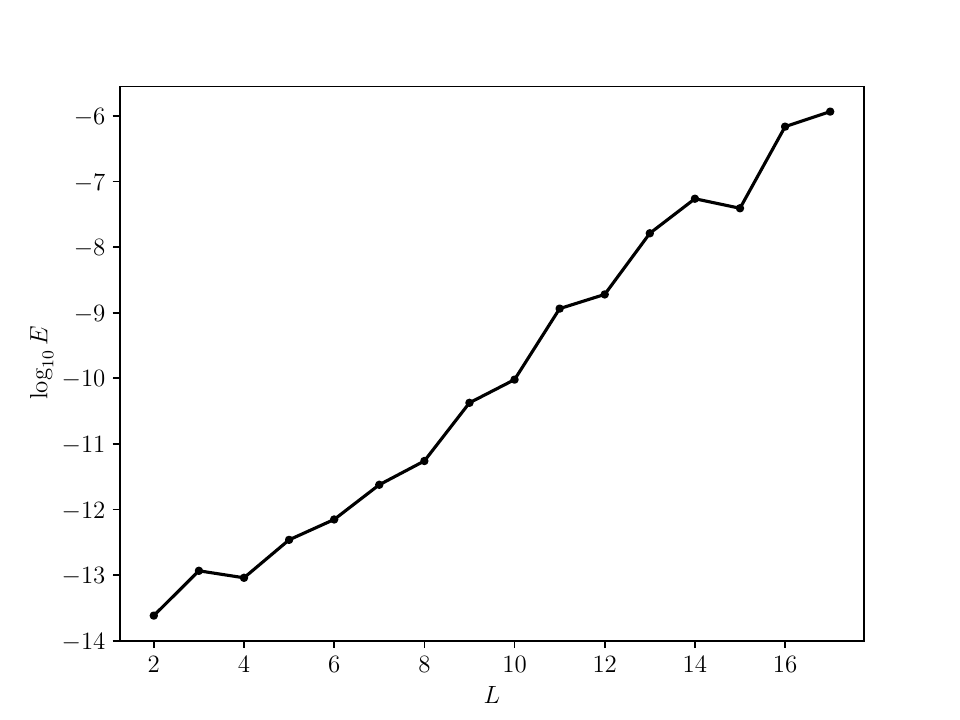}}
    \caption{(a) Error $E$ from Eq.~\eqref{eq:error_wg} with respect to the number $L$ of eigenvalues in the PCP and the modulus of the perturbation parameter $\nuv = \nuv_0 + \e^{0.3\i}\epsilon \begin{pmatrix}
    1\\
    1
\end{pmatrix}$. 
     (b) Evolution of the round-off error with respect to the number of eigenvalues in the set when $\epsilon=0$. Values obtained from hierarchical scheme with the initial guess $\nuv_0=(4.76715+7.01265\i, 2.470   +2.89872\i)$ and combining  2 to 19 eigenvalues with 5 derivatives in all directions.}\label{fig:error_duct}
  \end{center}
\end{figure}

Results are shown in Fig~\ref{fig:error_duct}a). It emerges that increasing the number of eigenvalues in the subset improves substantially the quality of reconstruction. The reason for this is that singularities due to the presence of exceptional points stemming from the coalescence of two (or more) eigenvalues belonging to the subset are removed.  When 10 eigenvalues are considered, the error remains below $10^{-1}$ in a large domain which covers almost the whole range of practical admittance values.
When eigenvalues are treated separately, i.e. without recombination, the truncated Taylor series of Eq.~\eqref{eq:lda_Taylor} (referred to as $\mathcal{T}_{\lambda}$) delivers poor results as expected, showing a very small domain of convergence limited by the presence of an exceptional point.
Curiously enough, it can also be observed that, when evaluated at the evaluation point ($\epsilon=0$), the best accuracy is obtained for low values of $L$. Though this seems to be contradictory with the above analysis, the loss of accuracy stems simply from the accumulation of round-off errors inherent in the PCP recombination process. This is well illustrated in Fig.~\ref{fig:error_duct}b) combining from 2 to 19 eigenvalues (with $D=5$ derivatives). 

\begin{figure}
  \begin{center}
   \subfigure[]{\includegraphics[width=0.48\textwidth]{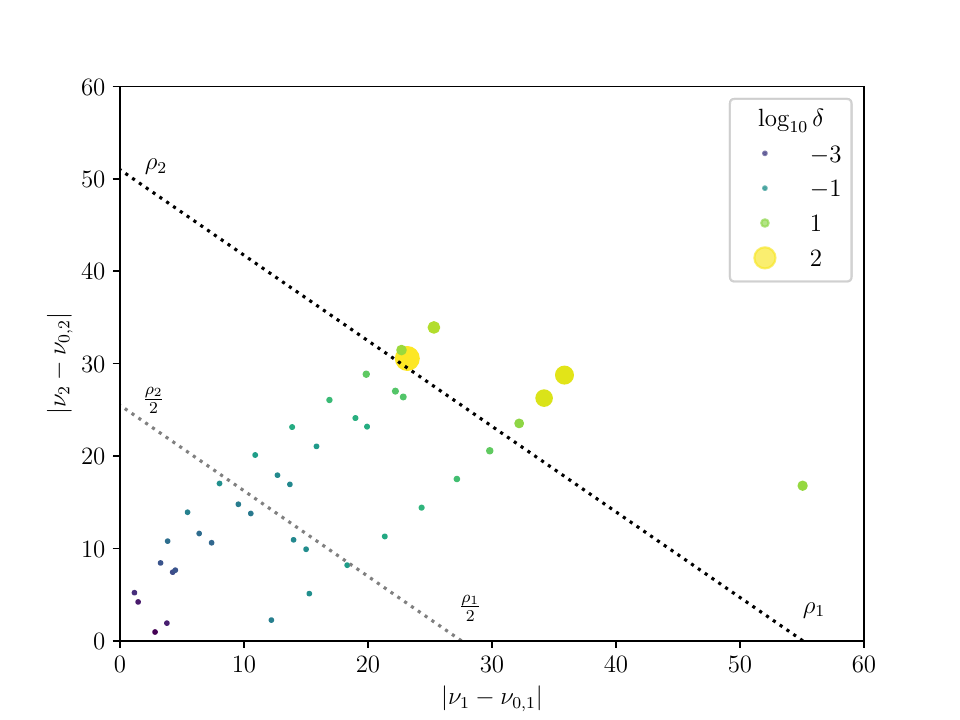}}  
   \subfigure[]{\includegraphics[width=0.48\textwidth]{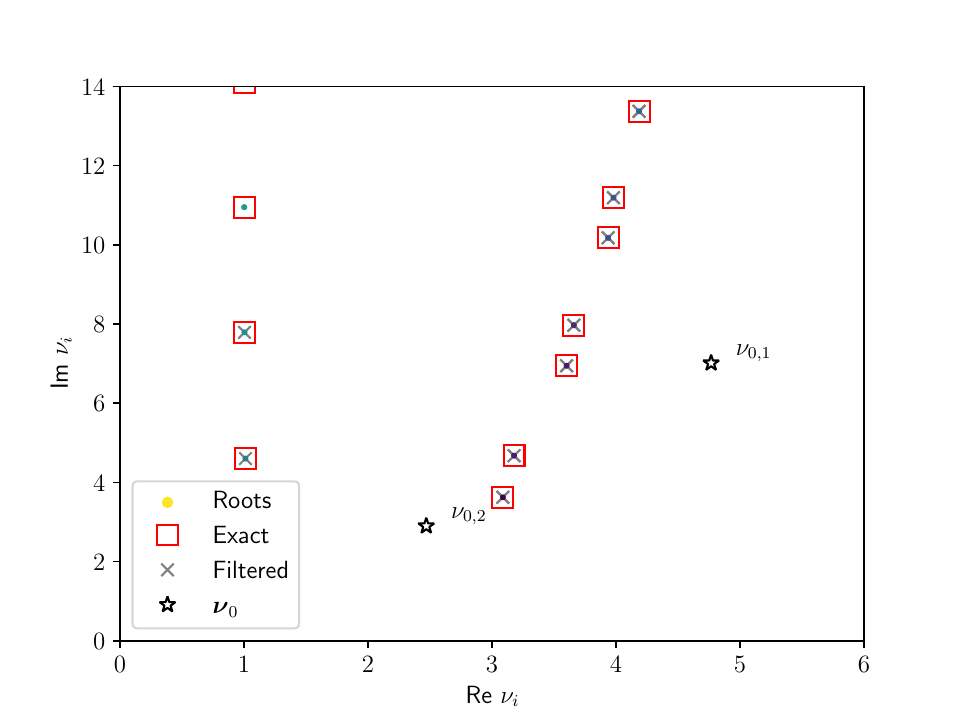}}
   \caption{(a) Distribution of $\mathrm{EP}_3$ solutions with their sensibility, computed with Eq.~\eqref{eq:sensitivity}, with respect to the parameters modulus $\abs{\nu_i - \nu_{0,i}}$ ($i=1, \, 2$). (b) $\mathrm{EP}_3$ solutions in the admittance complex plane after filtering  and comparison with reference solutions from \cite{Perrey-Debain:2021}. All solutions are obtained from a PCP obtained at $\nuv_0=(4.76715+7.01265\i, 2.470   +2.89872\i)$, combining 12 eigenvalues and with 5 derivatives in all directions. The root finding method uses the Levenberg-Marquard solver.}\label{fig:nu_plane_duct}
  \end{center}
\end{figure}

As opposed to the 3 degrees of freedom case (sec.~\ref{sec:toy-model}), there is an infinite number of $\mathrm{EP}_3$ for the continuous problem \eqref{eq:Helmholtz}, \eqref{eq:BC}  and \eqref{eq:Zwg}. These  correspond to specific values of the wall admittances associated with the coalescence of 3 duct acoustic modes \cite{perrey-debain2022}. 
In order to find some of these values from the discrete problem \eqref{eq:PVPz}, we choose a subset of 12 eigenvalues (whose magnitudes are ordered in ascending order from the lowest) and consider an order of derivation $D=5$ in both directions $\nu_1$ and $\nu_2$.
In order to use the iterative solver described earlier, all initial guesses $\sv_{\mathrm{i.\,g.}}$ are chosen from a regular grid $\mathcal{G}$ of points in the extended parameter space.
The grid covers a rather large domain since $\rho_i \approx 55$ ($i=1, \, 2$).
Now, results are conveniently shown in Fig.~\ref{fig:nu_plane_duct}a) where $\mathrm{EP}_3$ computed with the algorithm are plotted with respect to the distance from the evaluation point $\abs{\nu_i - \nu_{0,i}}$ ($i=1, \, 2$). 
This useful representation highlights the role of the radius of convergence. Each $\rho_i$ defines a parameter domain associated with solutions of sufficiently good quality. For instance, when $\abs{\nu_i - \nu_{0,i}} \leq \rho_i/2$ the  sensitivity indicator is below 0.1.
Now, using the filtering process with a threshold of $\delta =  2\cdot 10^{-2}$, the locations of some $\mathrm{EP}_3$ in the complex plane are shown in Fig.~\ref{fig:nu_plane_duct}b).
It is noteworthy that due to the symmetry of the waveguide, parameters $\nu_1$ and $\nu_2$ play a similar role and can be swapped.
These results show that more than 10 $\mathrm{EP}_3$ can be found with the same PCP.
Note that in all cases, the absolute error (exact solutions can be found in \cite{perrey-debain2022}) is found below $3\cdot 10^{-4}$ which proves the reliability of the filtering process.

\subsection{A 3D acoustic cavity problem}

\subsubsection{Problem statement}
\begin{figure}
    \centering
    \newcommand{\figheight}{3.5cm}
    \subfigure[$M=2,064$]{\includegraphics[height=\figheight]{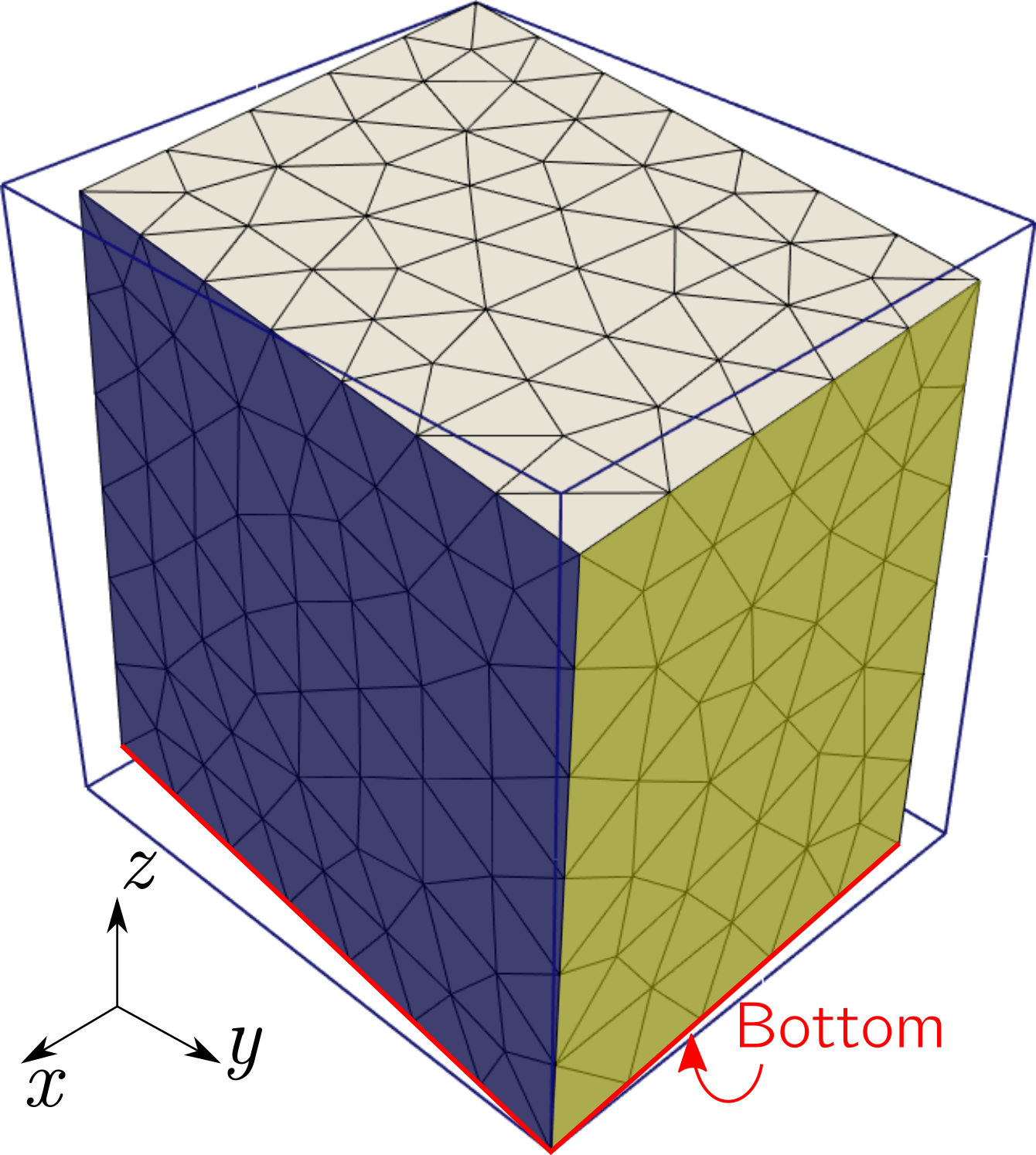}}   
    \subfigure[$M=76,557$]{\includegraphics[height=\figheight]{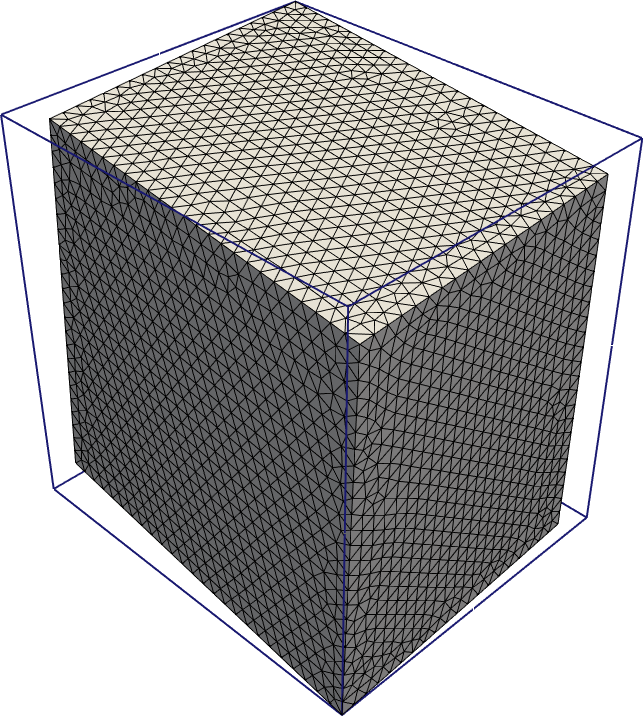}} 
    \subfigure[$M=326,725$]{\includegraphics[height=\figheight]{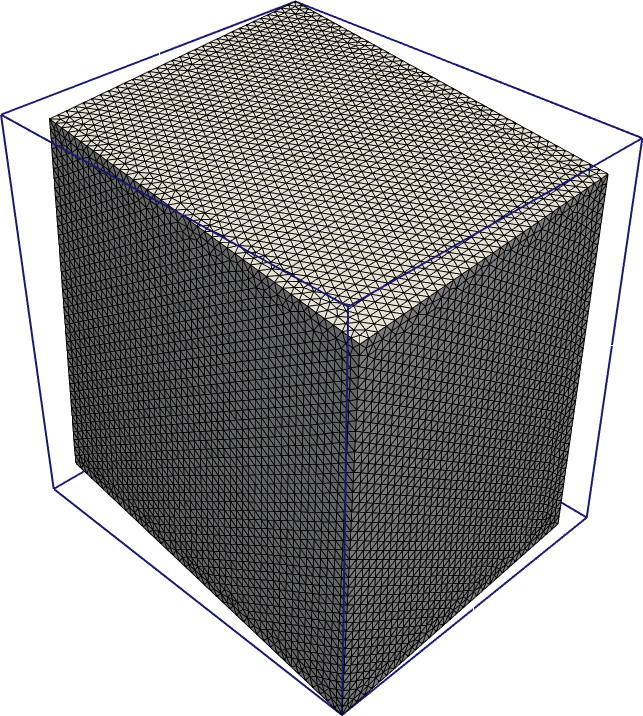}} 
    \subfigure[$M=555,668$]{\includegraphics[height=\figheight]{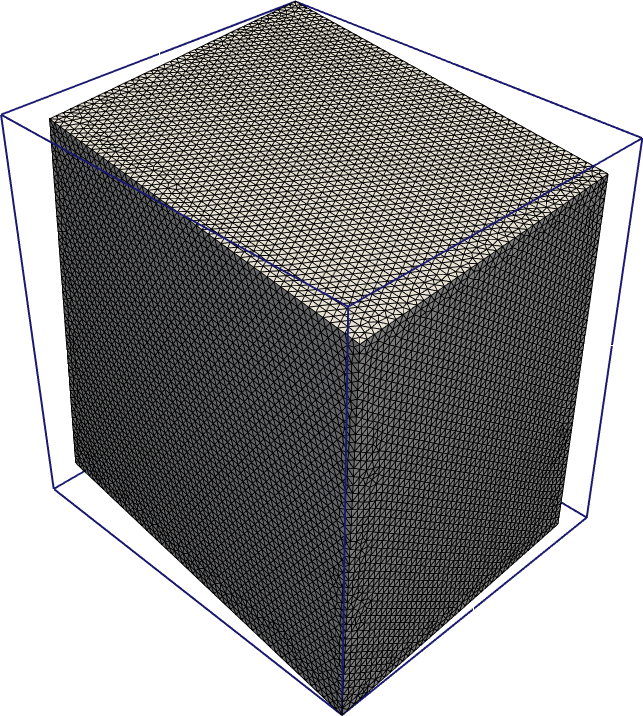}} 
    \subfigure[$M=1,054,708$]{\includegraphics[height=\figheight]{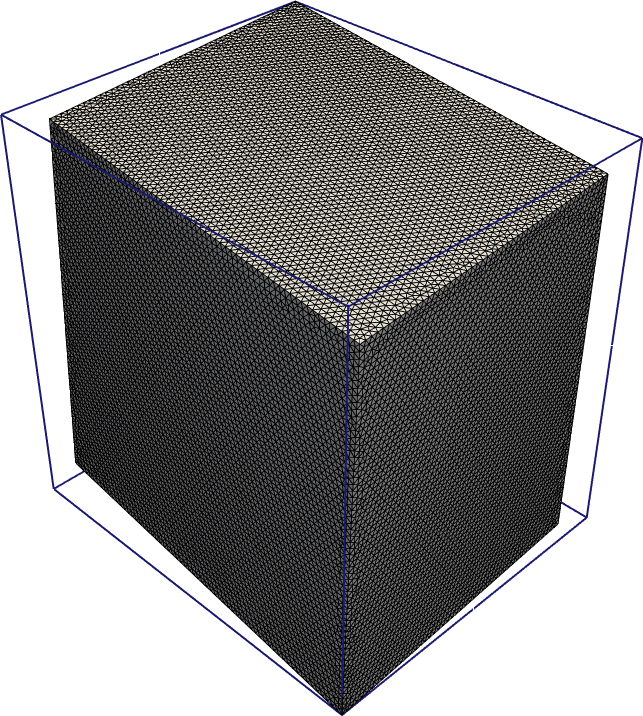}} 
    \caption{Meshes used in the 3D examples with the associated number of degrees of freedom $M$. The bounding box illustrates the deviation from a parallelepiped. Depending on the number of parameters in the tested configurations the first lined surface is in red (bottom), the second in blue and the third in green.}
    \label{fig:meshes}
\end{figure}

This last example represents a common scenario encountered in the field of noise control in an enclosed space where walls are treated with absorbing materials. For the sake of simplicity, the acoustic treatment at the walls is modeled assuming a local impedance boundary condition which is independent of frequency.
The existence of EP in this context has already been the subject of recent research work \cite{kanev2018}, in particular the link with the best modal decay rate has been shown.
In this numerical example, we are looking for acoustic modes of an enclosed cavity which has the shape of slightly deformed parallelepiped by using the following transformation
\begin{align}
    x &= 0.95x' + 0.05 y' - 0.07z', \nonumber\\
    y &= 0.01x' + 0.99  y' - 0.03333z', \\
    z &= -0.06x' - 0.08  y' + 1.01 z', \nonumber
\end{align}
where $(x',\,y',\,z')$ span the exact parallelepiped with dimensions $L_x=1$, $L_y=3^\frac{1}{5}$ and $L_z=2^\frac{1}{5}$.
This transformation has been chosen in order to ensure that the 6 faces $S_i$ ($i=1,\ldots,6$) remain flat but not parallel to each other, see Fig.~\ref{fig:meshes} (meshes are obtained with gmsh \cite{gmsh}). More importantly, it breaks the symmetry of the geometry which avoids some coincidental EP, the existence of semisimple eigenvalues and yields to a large number of EPs.
We introduce the parameters vector $\nuv=(\nu_1,  \nu_2, \nu_3)$ corresponding to the normalized wall admittance at the 3 walls ($i=1,2,3$):
\begin{equation}
    \nabla p \cdot \nv=  \nu_i p, \quad \mathrm{on} \, S_i,
\end{equation}
where $p$ is the acoustic pressure and $\nv$ stands for the unit outward normal vector.
The FEM discretization of the weak formulation associated to Helmholtz equation \eqref{eq:Helmholtz} is based on quadratic Lagrangian tetrahedral elements. This yields the following discrete eigenvalue problem (we put $\lambda=\kappa^2$ and $c=1$):
\begin{equation}\label{eq:PVPcav}
\Lv \big(\lambda({\nuv})), {\nuv}) \big) \xv({\nuv})  =  \big(-\Kv + \lambda({\nuv}) \Mv  +{\nu_1} \boldsymbol{\Gamma}_1 + {\nu_2} \boldsymbol{\Gamma}_2 
+ {\nu_3} \boldsymbol{\Gamma}_3 
\big)  \xv({\nuv}) =\mathbf{0},
\end{equation}
where $\Kv$ and $\Mv$ are the classical stiffness and mass matrices both real symmetric and (semi)positive definite. Matrices $\boldsymbol{\Gamma}_i$  stemming from the discretization of each wall are real-valued and symmetric. All FEM matrices are obtained from \cite{Dolfinx} and represented with PETSc \cite{Petsc} parallel matrices. Note that because the wall admittance $\nu_i$ is complex-valued, Eq.~\eqref{eq:PVPcav} defines a non-Hermitian system.
Here again, the application of the bordered matrix is simplified since only first derivatives of the operator are non-zero and
\begin{equation}
\partial_{\nu_i} \Lv= \boldsymbol{\Gamma}_i \text{  and  }
\partial_\lambda \Lv=\Mv.
\end{equation}

\subsubsection{Eigenvalue reconstruction quality}

\begin{figure}
\subfigure[$D=4$]{\includegraphics[width=0.18\textwidth]{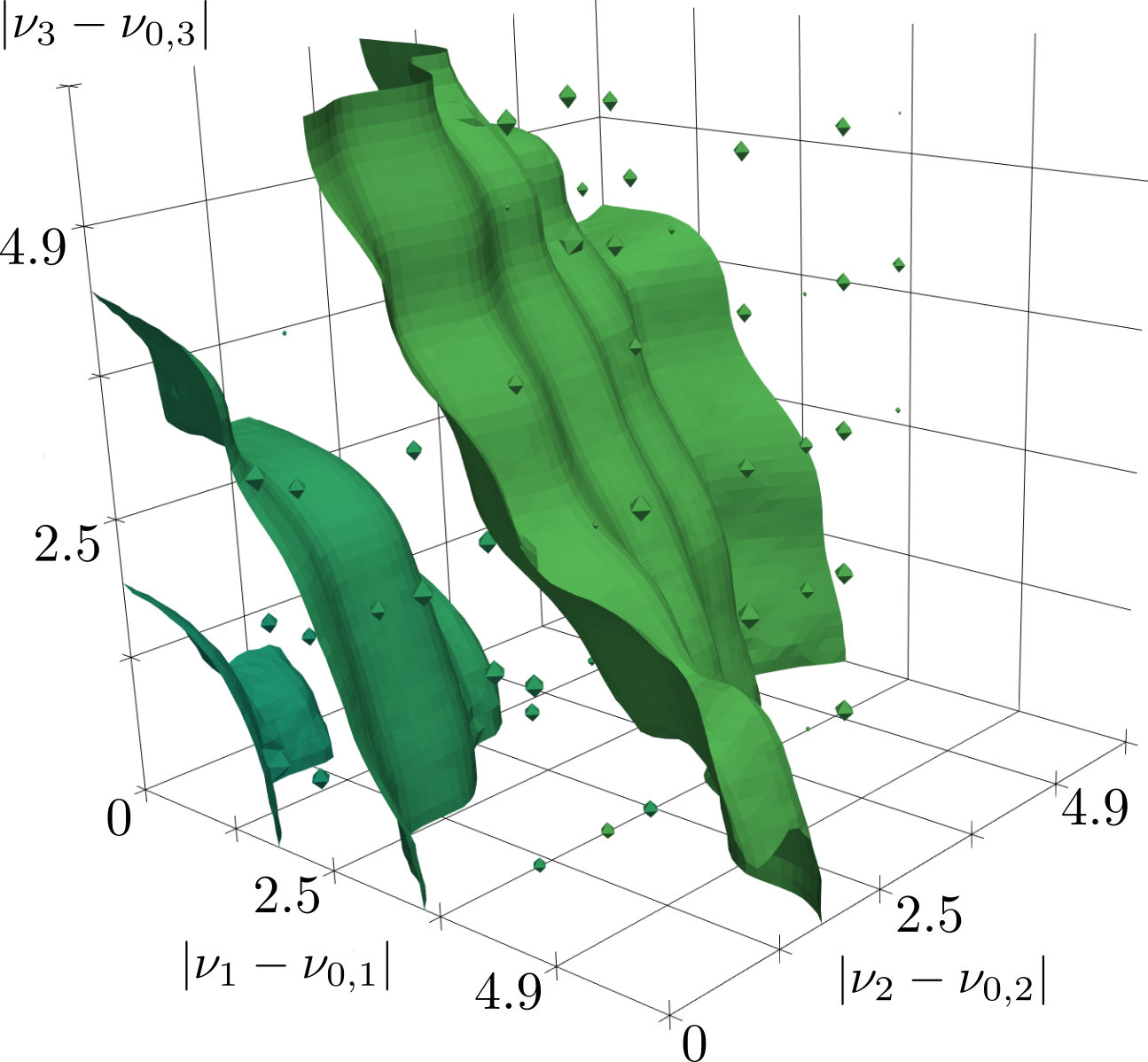}}
\subfigure[$D=6$]{\includegraphics[width=0.18\textwidth]{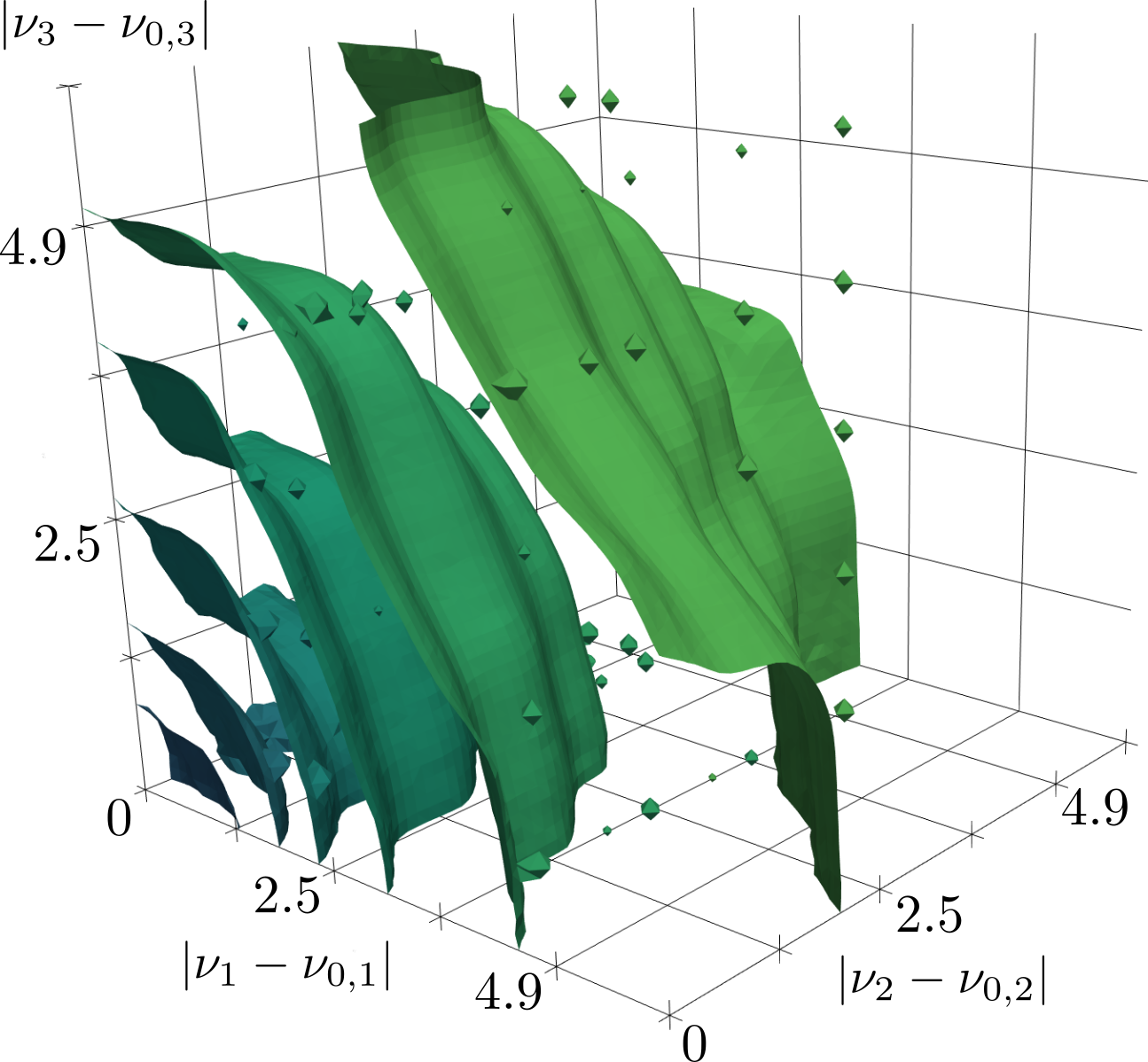}}
\subfigure[$D=8$]{\includegraphics[width=0.18\textwidth]{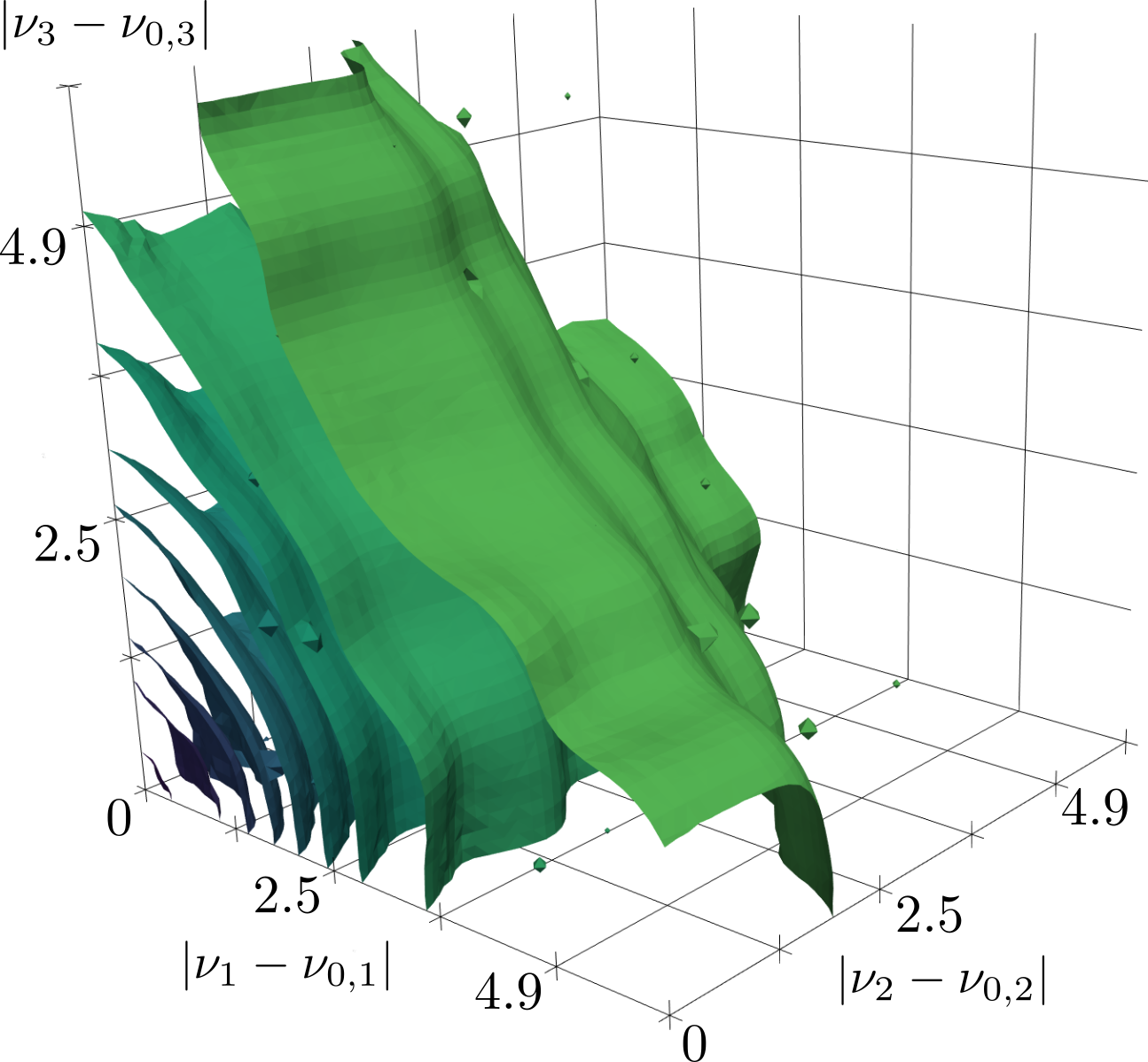}}
\subfigure[$D=10$]{\includegraphics[width=0.18\textwidth]{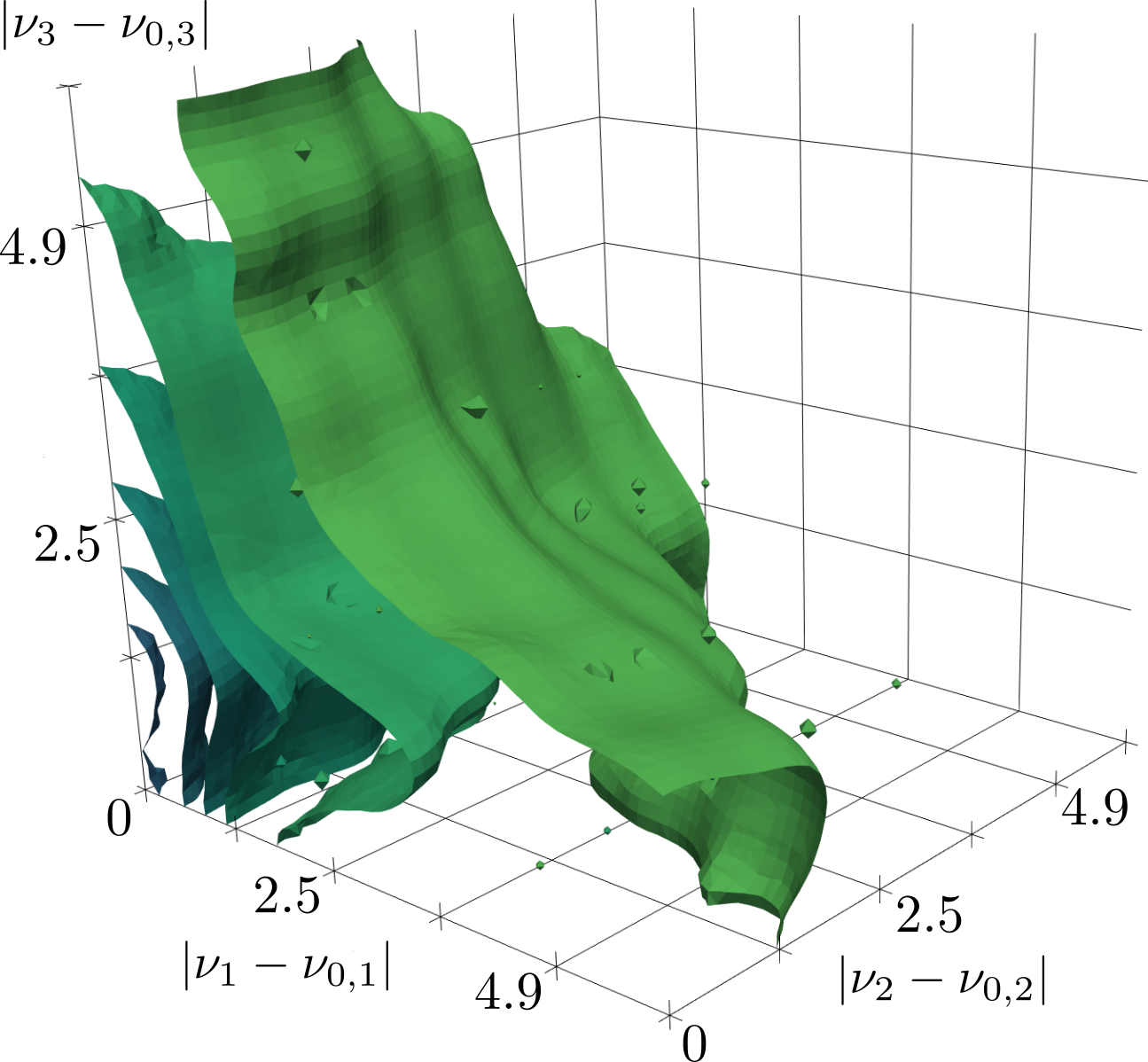}}
\subfigure[$D=12$]{\includegraphics[width=0.18\textwidth]{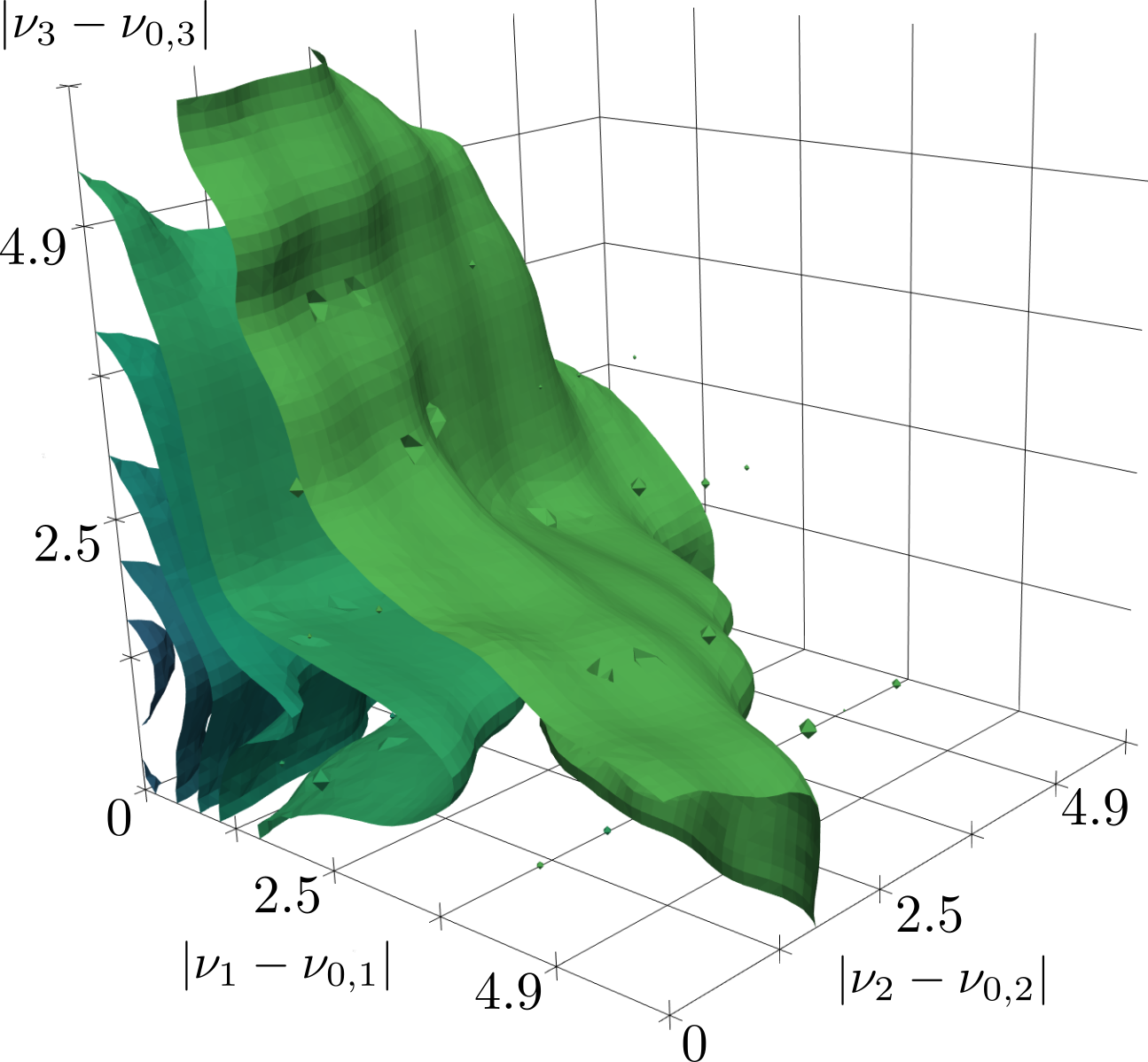}}
\subfigure{\includegraphics[height=0.17\textwidth]{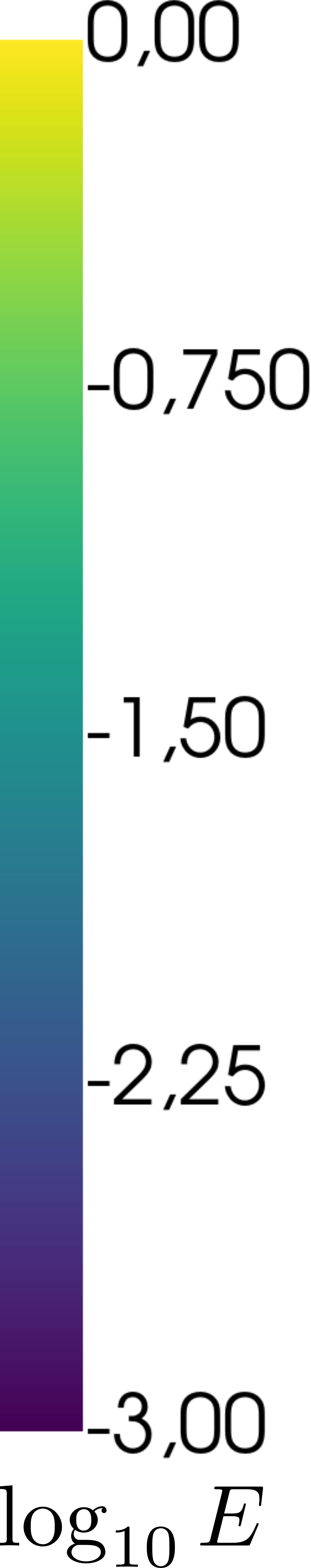}}
\caption{Iso-surface of the relative error between brute force direct computation and PCP evaluations with respect to the distance $\abs{\nu_i-\nu_{0,i}}$ in each direction $i=1,2,3$ and the derivative order. The dynamic is represented in log scale, dark blue correspond to $E=10^{-3}$ and the green to  $E=1$.}
\label{fig:bf_3p}
\end{figure}

In order to evaluate the accuracy of the eigenvalues  obtained from the PCP, reference solutions have been obtained via a ``brute-force'' approach, solving the original eigenvalue problem for each parametric values.
In total,  20  eigenvalues were computed  for a large number of parametric values around the evaluation point $\nuv_0 = (0.5-0.2\i, 1.2-1\i,  1 +0.5\i)$ and
\begin{equation}
    \nuv = \nuv_0 + (p_1 +\i q_1,p_2 +\i q_2,p_3 +\i q_3),
\end{equation}
where $p_i$ (resp. $q_i$) range from $-5$ to $5$ with eleven equispaced values. This yields $11^6 \approx 1.7\cdot 10^6$ eigenvalue problems to be solved and, in order to ease the computational burden, the coarsest mesh in Fig.~\ref{fig:meshes}a) is considered. 
On the other hand, the PCP is computed using a subset  $\Lambda$ containing only 11 eigenvalues. It is convenient for the analysis to define an error criterion as  the median 
\begin{equation}\label{eq:error_lda}
E(\nuv) = \mathrm{med}\, \lbrace\abs{\lambda_i^{\mathrm{pcp}}(\nuv)  - \lambda_i^{\mathrm{dir}}(\nuv) }\rbrace \,\, \text{for all $\lambda_i \in \Lambda$},
\end{equation}
where eigenvalues are ordered using the \emph{linear sum assignment} in order to find the best global match between eigenvalues obtained with the PCP and via direct computation.
Iso-values of the error are illustrated in Fig.~\ref{fig:bf_3p} for several values of the order of derivative $D$. For the need of representation, iso-values are plotted with respect to the 3 distances $|\nu_i - \nu_{0,i}|$.
Results show a very good quality of reconstruction in the vicinity of the evaluation point $\nuv_0$  when $D=8$ or $10$ whereas the case $D=6$ provides the best  overall approximation errors over a larger domain.
These numerical effects stemming from rounding errors show the subtle trade-off  between the number of eigenvalues retained in the subset and the truncation of the Taylor series.

\subsubsection{Higher order EP location}
\begin{figure}
\begin{center}
\subfigure[Face]{\includegraphics[height=0.45\textwidth]{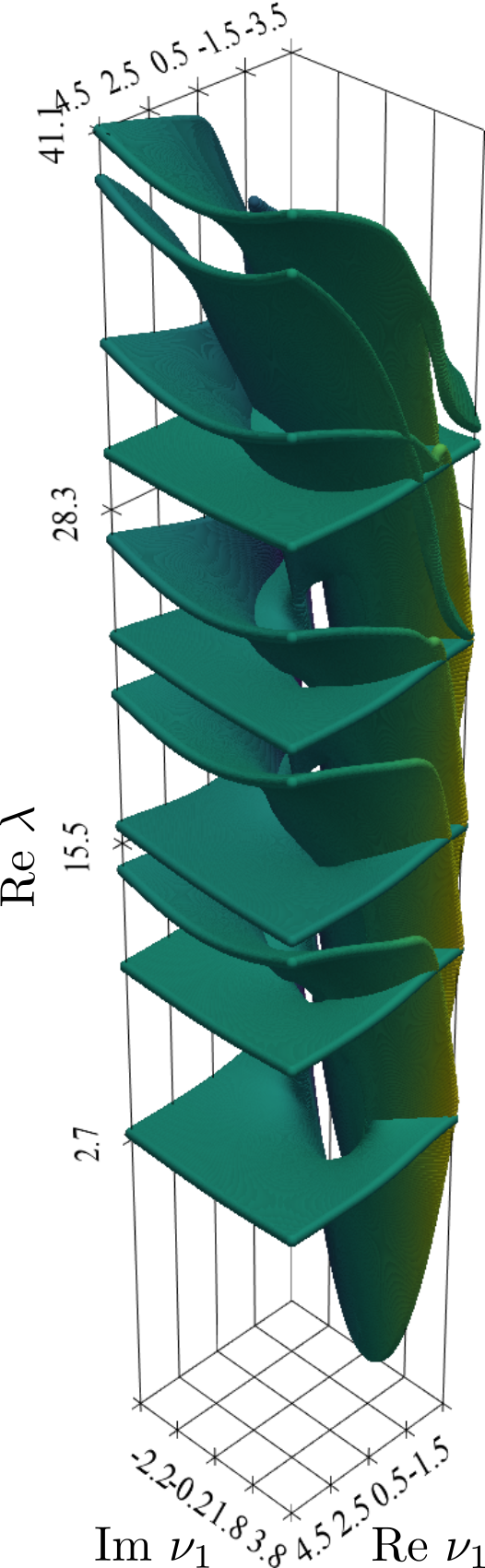}}
\subfigure[Back]{\includegraphics[height=0.45\textwidth]{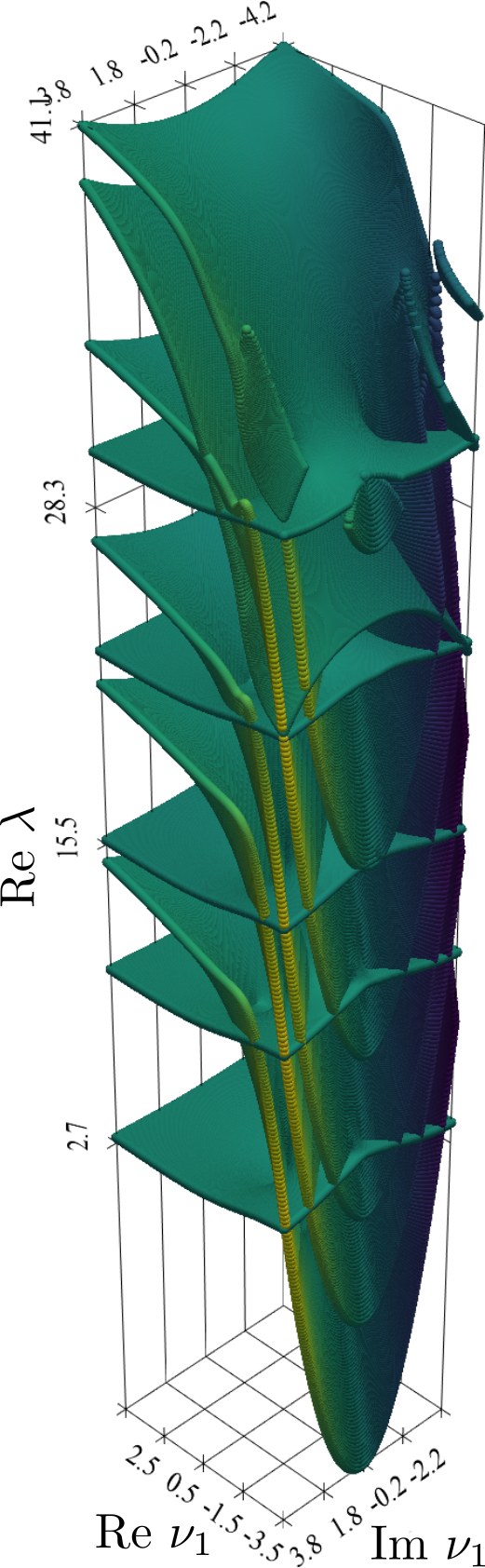}}
\subfigure[Discriminant]{\includegraphics[height=0.45\textwidth]{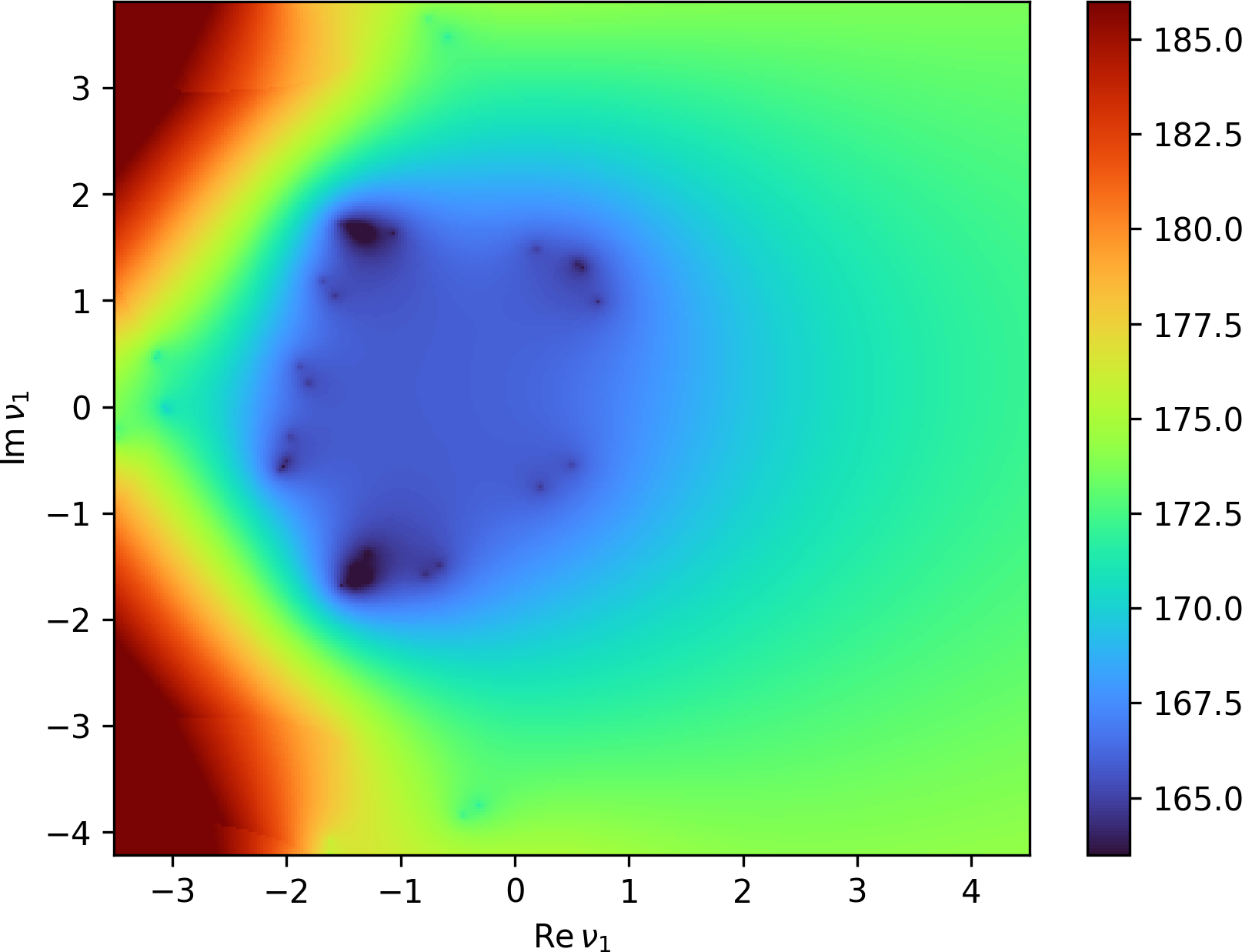}}
\end{center}
\caption{Direct computations of the Riemann surface face a), back view b) and the modulus  of the discriminant (log scale) c) around the evaluation point $\nuv_0$ for the first 11 eigenvalues (the imaginary part is depicted by the color).}
\label{fig:Riemman}
\end{figure}

In order to better illustrate the complexity of the approximation problem and understand the link with EPs, eigenvalues, obtained from direct computation, are plotted in Fig.~\ref{fig:Riemman} as function of a single complex-valued parameter $\nu_1$ while the other two parameters remain fixed to their original values. Because eigenvalues are complex-valued the color stands for the imaginary part.
This representation shows Riemann type surfaces which suggest the existence of exceptional points where eigenvalues coalesce. Their location is easier to observe using the discriminant 
\begin{equation}\label{eq:discriminant}
\mathcal{H}(\nuv) = \prod_{1\le i<j\le L} (\lambda_i(\nuv)-\lambda_j(\nuv))^2
\,\, \text{for all $\lambda_i \in \Lambda$},
\end{equation}
as shown in Fig.~\ref{fig:Riemman} c).
Here, dark spots suggest the existence of EPs in a certain neighborhood (ideally 
the coalescence of two (or more) eigenvalues should set the determinant to zero but numerical values are extremely sensitive in the vicinity of an EP).

For some admittance value, the resonant cavity problem  may tend to solution of lined semi-infinite duct that support surface modes. For 1D system, the problem is well established \cite{Pagneux:2013} and occurs when $\nu_1^2=-\lambda_n$. This leads to strong variation of the eigenvalues when $\mathrm{Re}\, \nu_1 < 0$ that favors crossings, thus EPs seem to be related to the transition between cavity and surface modes.

\begin{figure}
\begin{center}
\subfigure[]{\includegraphics[width=0.32\textwidth]{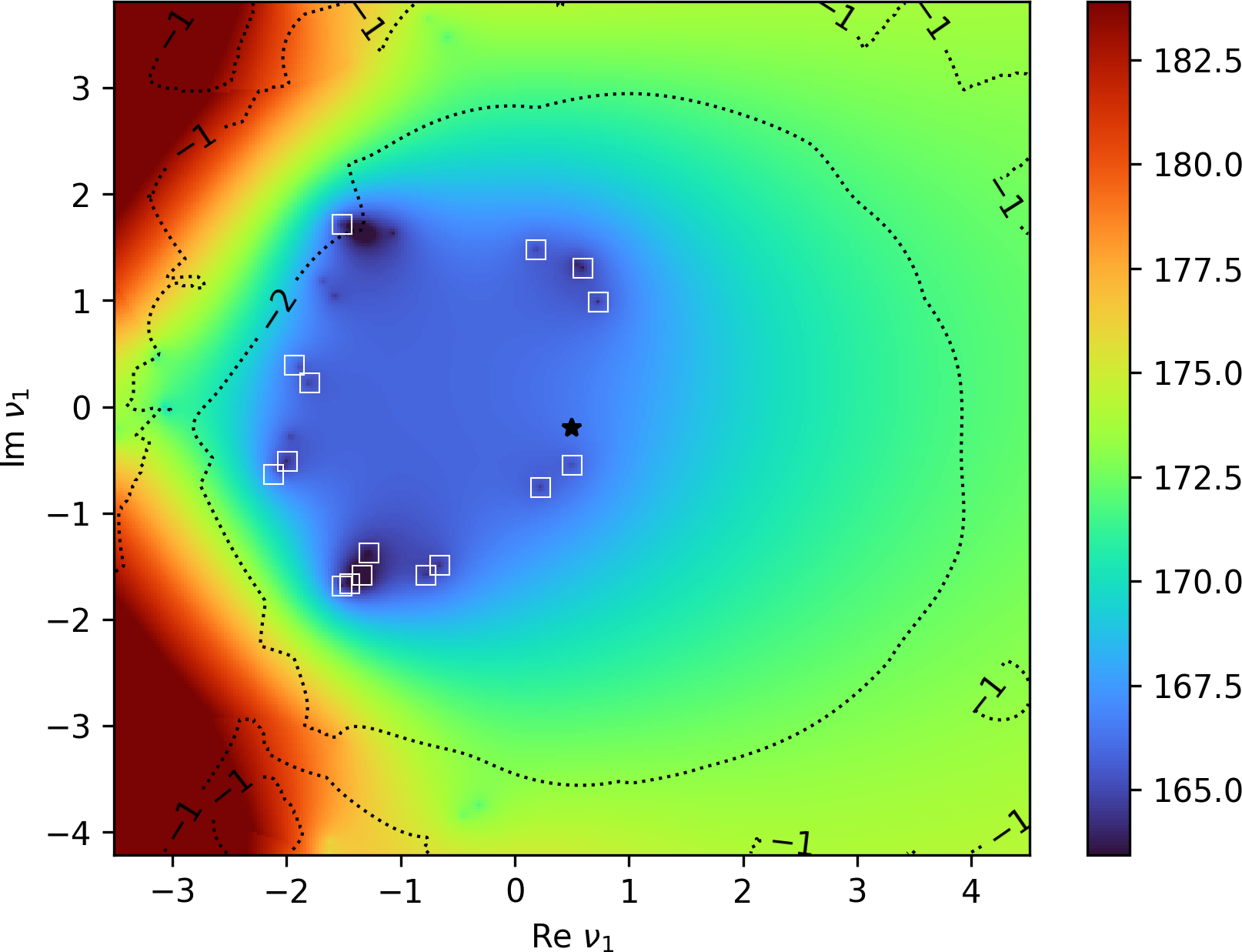}}
\subfigure[]{\includegraphics[width=0.32\textwidth]{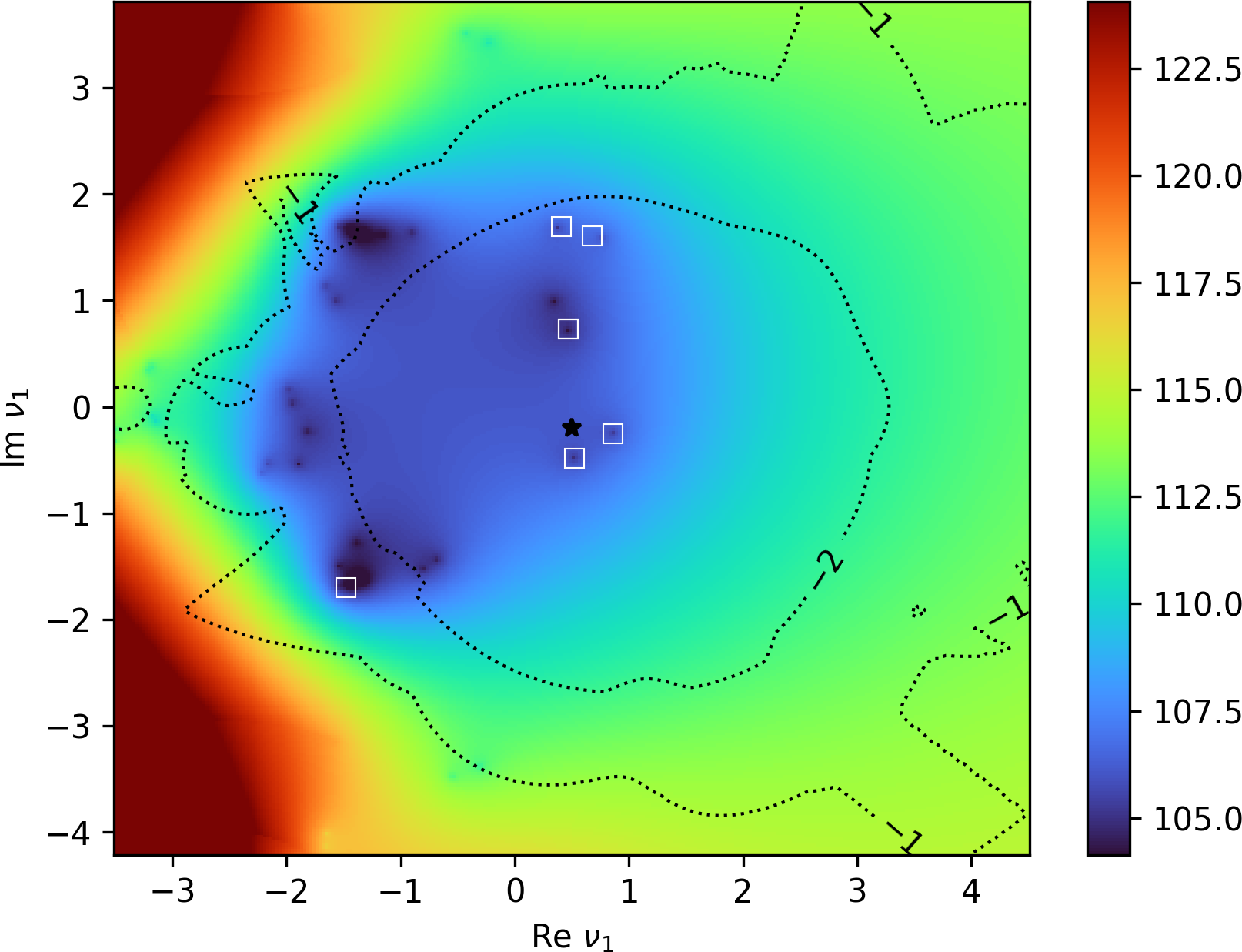}}
\subfigure[]{\includegraphics[width=0.32\textwidth]{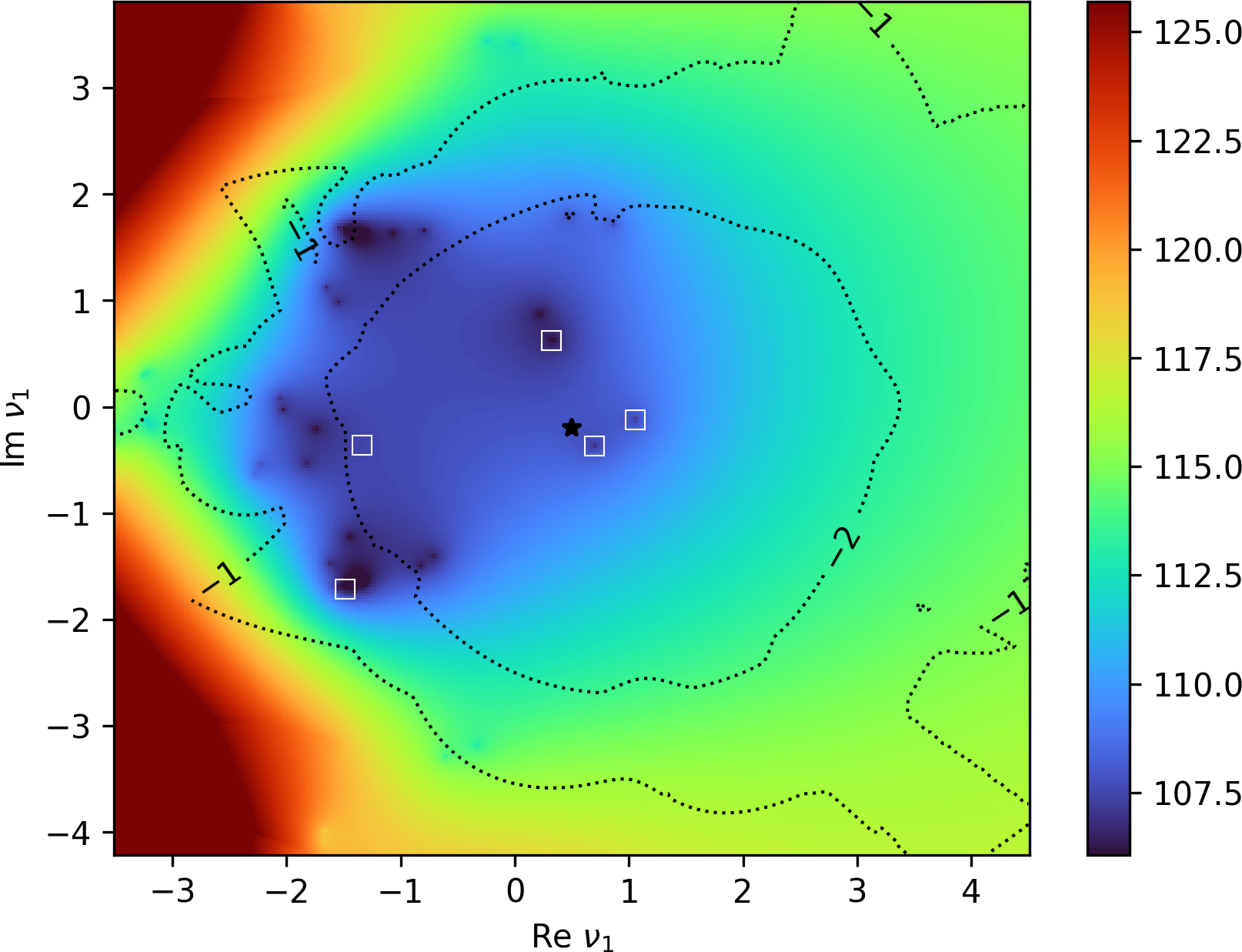}}
\end{center}
\caption{Exact discriminant with approximation error isoline at $10^{-1}$ and $10^{-2}$. \EP{2} found by the proposed algorithm are indicated by white square markers. In (a) $\nu_2=\nu_{0,2}$ and $\nu_3=\nu_{0,3}$ with $D=10$, in (b) $\nu_2=\nu_{0,2} + 0.353+0.353\i$ and $\nu_3=\nu_{0,3} + 0.353+0.353\i$ with $D=8$, in (c)  $\nu_2=\nu_{0,2} + 0.530+0.530\i$ and $\nu_3=\nu_{0,3} +0.530+0.530\i$ with $D=8$. The black star indicates the location of the evaluation point $\nu_{0,1}$.}
\label{fig:bf_1p}
\end{figure}

In order to assess the ability of the PCP for computing and retrieving the selected eigenvalues, other reference solutions (via direct computation) are computed and the associated discriminant is conveniently shown
with respect to $\nu_1$ in the vicinity of $\nu_{0,1}$ whereas other parameters are kept fixed, see Fig.~\ref{fig:bf_1p} a), b) and c).
Eigenvalues have also been recovered thanks to the PCP and error isolines are also displayed. 
This is somehow in agreement with reconstruction errors of Fig.~\ref{fig:bf_3p}. %
It emerges that eigenvalues can be recovered form the PCP with at least two digits of accuracy as long as  $|\nu_1 - \nu_{0,1}|\lesssim  2 $ and this is in line with the theoretically predicted radii of convergence $\rho_i$ ($i=1,\,2,\,3$), defined from Eq.~\eqref{eq:roottest} which is found to be around 7.  Based on the authors' experience with this particular point, it is found that $\rho_i/2$ usually provides a reliable upper bound for the PCP. The interest for the PCP from which eigenvalues are computed at negligible cost is clearly shown here even in the presence of a relatively large number of EPs. For comparison, the eigenvalue Taylor series have radii of convergence between 0.6 and 4.5.

Figs.~\ref{fig:bf_1p} also indicates the location of \EP{2} calculated with our algorithm and identified with white square markers (it is unlikely to spot higher order EPs in this colormap since only one varying parameter, i.e. $\nu_1$ is considered). 
Here , \EP{2} solutions have been selected with a sensitivity indicator (Eq.~\eqref{eq:sensitivity}) ranging between 0.2 and $10^{-4}$.
When the fixed parameter values are shifted from their original settings, the radius of convergence $\rho_1$ decreases slightly, as well as the number of \EP{2} found in the region of interest, as shown in Fig.~\ref{fig:bf_1p} b) and c).

If we now fix only one parameter $\nu_3=\nu_{0,3}$ (with the same evaluation point $\nuv_0$), the parameter space belongs to $\mathbb{C}^2$ and third order EP can be found using the same algorithm. In practice however, it is found  that the sensitivity indicator is not as good especially when the  $\mathrm{EP}_3$ lies close to the boundary of the convergence region (say further than $\rho_i/2$). This is the case in the example presented here with a sensitivity of about 0.1. A refinement step is thus necessary in order to achieve better accuracy. It suffices for that to consider the value just found as a new evaluation point, here we take:
$\nuv_0' = (0.46-0.22\i, -1.08-1.82\i) + 0.2\i(1,\, 1)$ 
where $0.2\i$ is a small arbitrary quantity which is  added to avoid a nearly defective matrix system (this would otherwise have a detrimental effect on the eigenvalue solver and the computation of derivatives).
A new PCP can now be built with the same $\lambda$ set and the same derivative order. 
Nonetheless, the proximity to the exact location means that a smaller order of derivation could be exploited to limit the numerical cost of the process. In the same way, a smaller subset $\Lambda$ containing exclusively the three eigenvalues which are known to coalesce could be retained.

\begin{figure}
\centering
\includegraphics[height=4cm]{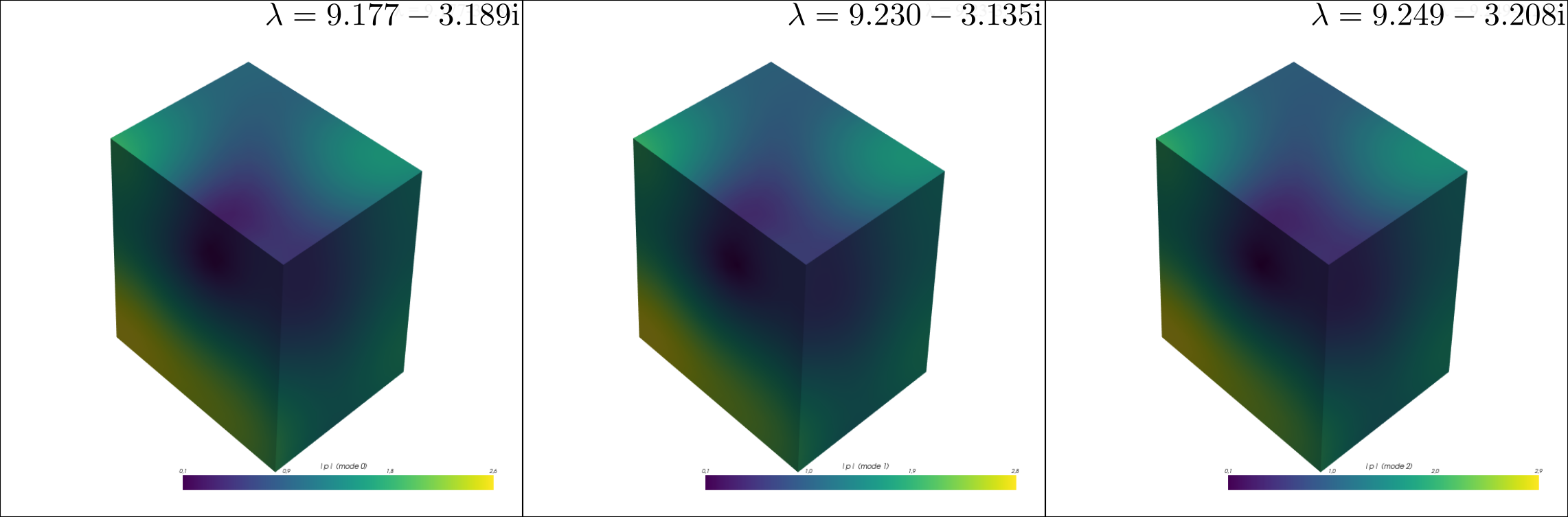}
\caption{Amplitude of the pressure field for the three nearly-coalescing eigenvectors obtained with $\nu^*_1=0.36931-0.06183\i$, $\nu^*_2=-1.07538-1.84876\i$ and $\nu_3=1+0.5\i$ (eigenvalues are shown in the inset). }\label{fig:ep3_field}
\end{figure}

The new solution is found to be $\lambda^*=9.21872-3.17739\i $ with
\begin{equation*}
\nu^*_1=0.36931-0.06183\i \quad \mathrm{and} \quad \nu^*_2=-1.07538-1.84876\i,
\end{equation*}
whereas $\nu_3=\nu_{0,3}=1+0.5\i$ is fixed. The quality of the solution is guaranteed by the sensitivity indicator which is below $10^{-4}$.
The pressure field, obtained from direct computation with this set of parameters, can be seen in Fig.~\ref{fig:ep3_field}. It is clear that the three eigenvectors are nearly identical and eigenvalues differ from each other by less than  1\%. It is noteworthy that $\lambda^*$ value is nearly the mean of the three computed eigenvalues which was expected from the Puiseux series.

\begin{figure}
\centering
\includegraphics[height=4.5cm]{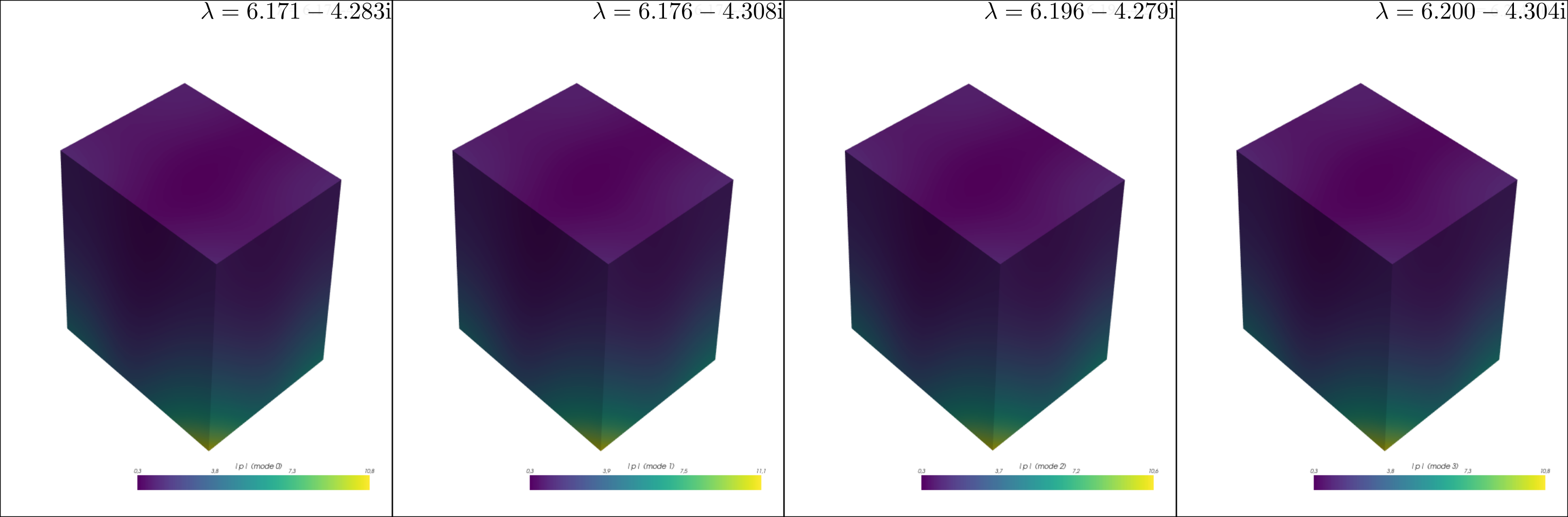}
\caption{Amplitude of the pressure field for the four nearly-coalescing eigenvectors obtained with  $\nu^*_1=-1.76495-0.86480\i$, $\nu^*_2=-1.24735-2.10971\i$ and $\nu^*_3=-0.78546+1.80176\i$ (eigenvalues are shown in the inset). Here $D=8$ and $L=8$.}\label{fig:ep4_field}
\end{figure}

By now letting the 3 parameters free, \EP{4} should exist even though things are more tricky from a computational point of view as explained earlier. However, the same refinement process can be carried out. 
Eventually, by taking $\nuv_0''=(-1.44-0.38\i, -1.35-1.72\i, -0.44+1.33\i)$ as the evaluation point and a solution is found to be $\lambda^*=6.18579-4.29223\i $ with
\begin{equation*}
\nu^*_1=-1.76495-0.86480\i,  \quad \nu^*_2=-1.24735-2.10971\i
\quad \mathrm{and} \quad \nu^*_3=-0.78546+1.80176\i.
\end{equation*}
The pressure field, obtained from direct computation with this set of parameters can be shown in Fig.~\ref{fig:ep4_field}. It is clear that the four eigenvectors are nearly identical and eigenvalues differ from each other by less than  0.5\%.

We end this section by noting that if high order EPs are more difficult to identify due to the extreme sensitivity\footnote{It can be shown that $|\lambda - \lambda^*| = \mathcal{O}(\varepsilon^{1/N})$
given the parametrization $\nuv= \nuv(\varepsilon) \in \mathbb{C}^N$ and $\nuv(0)=\nuv^*$.} in the vicinity of the EP
the algorithm proposed in this work, along with the refinement steps as described above, proved to be a reliable tool for identifying high order EPs in a parameter space.

\subsubsection{Complexity and computation time}\label{sec:time}
\begin{table}
\begin{center}
\begin{tabular}{rcccccc}
\hline
$M\,(\#\mathrm{CPU})$ & Eig. & LU & FB sub. & Built RHS & \multicolumn{2}{c}{Total time for $12\times 12$ Derivatives } \\
\cline{6-7}
- & s & s & s & s & s & /Eig.  \\
\hline
2,064~~(1) & 0.19 & 0.023  & 0.00 & [0.01, 0.11] & 1.39 & 7.36 \\
2,064~~(2) & 0.19 & 0.020  & 0.00 & [0.01, 0.11] & 1.23 & 6.54 \\
2,064~~(4) & 0.15 & 0.020  & 0.00 & [0.01, 0.10] & 1.12 & 7.65 \\
76,557~~(4) & 9.2 & 3.0  & 0.05 & [0.06, 0.28] & 18.2 & 1.98 \\ %
76,557~~(8) & 6.1 & 2.2  & 0.03 & [0.04, 0.21] & 12.2 & 1.99 \\
76,557~(12) & 5.2 & 2.0  & 0.03 & [0.03, 0.19] & 10.8 & 2.07 \\
326,725~~(4) & 83 & 34  & 0.46 & [0.28, 1.00] & 140 & 1.70 \\
326,725~~(8) & 43 & 22  & 0.23 & [0.15, 0.59] & 76 & 1.78 \\
326,725~(12) & 38 & 18  & 0.21 & [0.12, 0.47] & 64 & 1.70 \\
555,668~~(8) & 100 & 56  & 0.46 & [0.27, 0.98] & 161 & 1.61 \\
555,668~(12) & 85 & 46  & 0.41 & [0.21, 0.76] & 134 & 1.57 \\
555,668~(16) & 80 & 41  & 0.39 & [0.18, 0.69] & 123 & 1.54 \\
1,054,708~~(8) & 313 & 184  & 1.23 & [0.66, 2.33] & 455 & 1.45 \\
1,054,708~(12) & 258 & 155  & 1.09 & [0.50, 1.84] & 382 & 1.48 \\
1,054,708~(16) & 225 & 126  & 0.95 & [0.42, 1.48] & 322 & 1.43 \\
\hline
\end{tabular}
\caption{CPU time for the main computational steps for several problem sizes (see meshes in Fig.~\ref{fig:meshes}). The computational work is carried out using parallel computing on multiple processors (1, 4, 8, 12 and 16). The `build RHS' time corresponds to the mean value and the max value respectively. Computations are performed on Intel(R) Xeon(R) Silver 4309Y CPU \@ 2.80GHz with 256 Go of RAM. The `Total time' column includes 144 RHS builds and FB in addition to the initial LU factorization.  (Global RHS building time can be obtained by multiplying the mean value by $12^2$).}
\label{tab:time}
\end{center}
\end{table}

This last section exclusively focuses on the assessment of the computational cost of the proposed method. This discussion aims to measure the computational gain compared to direct approach on the original algebraic system. Results were presented in \cite{Nennig:2020} for $N=1$. The discussion here focuses on the effects of the number of parameters $N$ and the number of derivative $D$.

Tab.~\ref{tab:time} reports the computation time needed to compute the eigenvalue Taylor coefficients up to $D=12$. Here, only a single eigenvalue depending on two parameters ($N=2)$ is considered. In this case, it can be observed that between 20\% and 40\% of the global time is impacted by the LU factorization of the bordered matrix which is required for the computation of the first derivative. Higher derivatives ($12 \times 12 =144$ in total) are comparatively cheaper to compute thanks to the efficiency of the forward and backward (FB) substitution used to solve \eqref{eq:Bordered} for the multiple RHS.%

The last column in Tab.~\ref{tab:time} shows the ratio between the global cost, indicated by the total time, and the computation time for one run of the eigenvalue solver. If the ratio is relatively high for small size matrices for which the eigenvalue computation is cheap, it tends to 1 for large algebraic systems. In this situation, the cost for the eigenvalue derivatives becomes marginal and this renders the method very advantageous for large size matrices or for more complicated problems like quadratic eigenvalue problems. To give some element of comparison, if finite differences were used, more than five eigenvalue problem solves are already required to compute all partial derivatives up to the second order. %

The algorithm can benefit from parallelism through eigenvalue computation \cite{SLEPc}, linear solver \cite{Mumps, Petsc} and to build the numerous matrix/vector multiplications from RHS (see \eqref{eq:Bordered}). The peak memory usage arises during LU factorization.

It is worth noting that the number of terms appearing in each RHS and the number of RHS depends strongly on the order of the derivatives.
The crossover point between the linear solve and the construction of the RHS building is reached when $D=9$. If $D$ is further increased, the time for the computation of the RHS becomes dominant as illustrated in Tab.~\ref{tab:time}.

\begin{figure}
    \centering
    \includegraphics[width=0.5\textwidth]{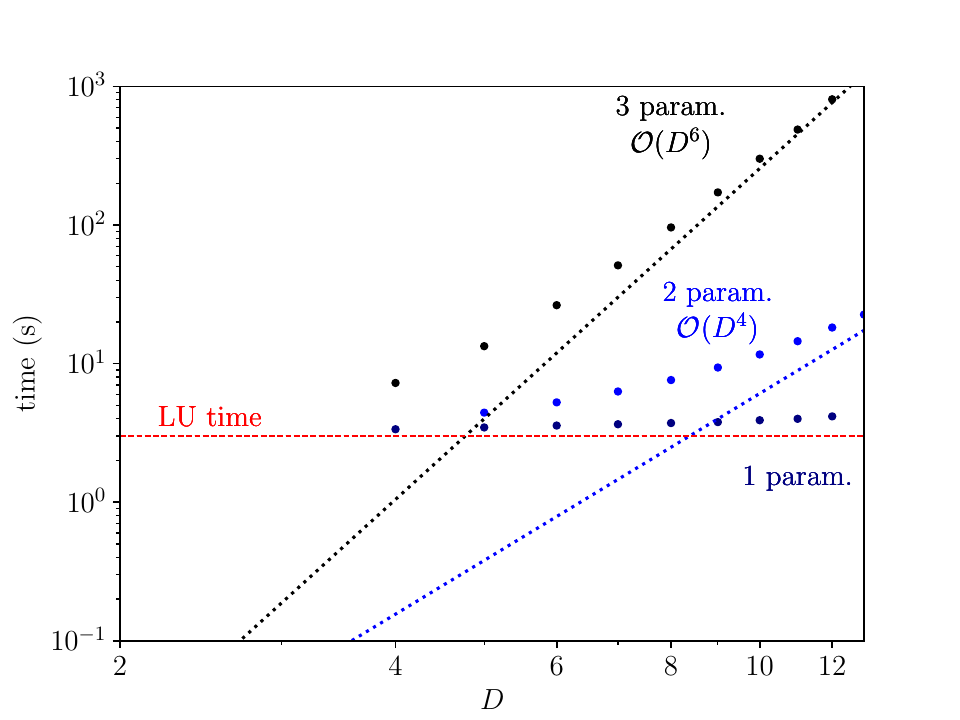}
    \caption{Total derivative time with respect to the number of derivatives $D$ for each parameter. The number of parameter may be 1 (dark blue), 2 (blue) or 3 (black). The dotted lines indicate the theoretical asymptotic complexity and the red dashed line represent the inital LU factoraization step.}
    \label{fig:RHS}
\end{figure}

\begin{table}
\begin{center}
\begin{tabular}{rrcccccc}
\hline
$N$ & $D$ & Built RHS & \multicolumn{2}{c}{Total derivatives time } & &\multicolumn{2}{c}{PCP}\\
\cline{4-5}\cline{7-8}
- & - & s & s & /Eig. && s (seq.) & s (4) \\
\hline
2 & 6 & [0.01, 0.05] & 5 & 0.57 && 0.03& \\
~ & 8 & [0.03, 0.09] & 8 & 0.90 && 0.09& \\
~ & 10 & [0.04, 0.17] & 12 & 1.39 && 0.18& \\
~ & 12 & [0.06, 0.28] & 18 & 1.98 && 0.43& 0.32\\
~ & 16 & [0.11, 0.53] & 42 & 5.45 && 1.18& 0.55\\[0.8em]
\hline
3 & 4 & [0.02, 0.13] & 7 & 0.82 && 0.09& \\
~ & 6 & [0.06, 0.36] & 26 & 3.40 && 0.58& 0.32\\
~ & 8 & [0.14, 1.31] & 96 & 12.12 && 2.92& 1.46\\
~ & 10 & [0.25, 4.61] & 301 & 32.49 && 10.02& 3.85\\ %
~ & 12 & [0.42, 11.66] & 807 & 87.76 && 27.81& 10.78\\
\hline
\end{tabular}
\caption{CPU time for the main computational steps for different derivation order and number of parameter. Computations are performed on Intel(R) Xeon(R) Silver 4309Y CPU \@ 2.80GHz with 256 Go of RAM. All computation are performed on the mesh in Fig.~\ref{fig:meshes}(b) with 76k dof using 4 CPU. For the normalization, 10.2 s are used for the eigenvalue computation time (see Tab.~\ref{tab:time}). The PCP times is obtained for 11 eigenvalues in $\Lambda$ in sequential (seq.) or with 4 CPU.}
\label{tab:timeD}
\end{center}
\end{table}
The number of RHS always grows as $\mathcal{O}(D^{N})$. For this problem, we can also find an expression for the cost of each RHS computation when exploiting the fact that the stiffness matrix depends linearly on the parameters vector $\nuv$. %
It can be seen from the general formulas \eqref{eq:Leibniz_code} and \eqref{eq:Leibniz_codeK} that (i) there are $N$ terms in the RHS vector stemming from the stiffness matrix and (ii) there is a cumulative sum of $\alpha$ ($\alpha=1,\dots, D)$ terms to compute the derivative up to $\alpha$ for each of the $N$ variables stemming from the mass matrix.
One can estimate that the computation time leading order needed to build the RHS must scale like $\mathcal{O}(D^{2 N})$.
In the situation where $N=2$ parameters are used (see Tab.~\ref{tab:timeD}), it can be seen that this time grows like $\mathcal{O}(D^{4})$. Similarly, if $N=3$ the RHS computation time grows like $\mathcal{O}(D^{6})$ and the computation for the RHS becomes dominant for $D\ge 5$ over the initial LU solve. These predictions are best illustrated in Fig.~\ref{fig:RHS}.
For small values of $D$, the total derivatives time correspond to the LU factorization, but higher value of $D$ the RHS asymptotics are reached.

If the number of parameters becomes too large (say above 5), our experience shows that the proposed approach find its limitation (at least for high derivative order) although there is room for improvements for instance by using  parallel computing for the derivation process and for the construction of the RHS vector.
To nuance this remarks, when the number of degrees of freedom is increased (at fixed $N$ and $D$), the RHS building time scales as $\mathcal{O}(M)$ whereas both the linear solver and the eigenvalue solver contribute significantly to the overall computational burden. %

Similar observations can be made for the computation of the PCP. If its computation is negligible for one or two parameters, it can be significant above three parameters especially if more than six derivatives are used in each direction. As illustrated in the last column of Tab.~\ref{tab:timeD}, it is possible to take advantage of parallel computing to speed up the summation in \eqref{eq:Leibniz_prod_mv}. Note that nested scalar operations in Python are used in this work but improvements are possible by using code optimization using compiled codes.

\section{Conclusion}

In this paper, a numerical algorithm dealing with parametric non-Hermitian eigenvalue problems has been presented.
The main idea relies on the concept of a Partial Characteristic Polynomial (PCP) whose coefficients are regular functions in a large domain of the parameter space and from which a subset of selected eigenvalues can be straightforwardly recovered at a negligible cost. The PCP can also be used to find high order Exceptional Points (EP) which have raised considerable attention in the scientific community, especially in physics but also in mechanics. 
EPs are special degeneracies corresponding to specific values of the system parameters for
which both eigenfrequencies and eigenvectors coalesce simultaneously. 

A large part of the paper is devoted to a detailed presentation  of all the theoretical and numerical ingredients which are needed for the different building blocks of the algorithm: the construction of the PCP is the subject of Sec.~2 while  Sec.~3 presents different numerical strategies for the location of EPs from the knowledge of the PCP.
Through numerous numerical tests, three examples of increasing complexity are presented in the last section showing the numerical stability and the computational efficiency of the method. In this respect, it is shown that the method offers a substantial saving of computational resources if one is interested in the parametric study of a selected subset of eigenvalues associated with  large
size sparse matrices arising from conventional discretization techniques such as the finite element method.

The main governing parameters of the proposed method are (i) the number of eigenvalues retained in the subset, (ii) the number of terms in the truncated Taylor series and (iii) the number of complex-valued parameters considered in the study. %
These parameters, when they become too important, can lead both to the accumulation of round-off errors and to the computational burden caused by the construction of the right-hand side vector.
In practice, it is found that the number of eigenvalues $L$, the order of derivation $D$ and the number of parameters $N$ must be kept within reasonable values 
(say $2<L<18$, $D<16$, $N<5$). However, the method is applicable to large (usually sparse) matrices arising from FE discretization and this renders the approach appealing for the study of real-life applications.
The method presented in this work is, to the best of the author's knowledge, quite novel and there are good reasons to believe that there is room for improvement. Numerical examples presented in the paper were chosen mainly to illustrate the method. 
One particular industrial application relates to the application of this technique for optimal design of acoustic treatment. In the context of acoustic duct, the acoustic treatment at the wall is known to be associated with the existence of EPs. This application would be the object of further investigations, exploiting the benefits of the proposed approach.

\appendix
\section{Eigenvalue partial derivatives computation}\label{app:derivative}

Here, the successive partial derivatives $\partial^{\alphav}\lambda$ of each $\lambda\in \Lambda$ and of its associated eigenvector $\partial^{\alphav} \xv$, satisfying
\begin{subequations}
\begin{align}
\Lv (\lambda(\nuv), \nuv) \xv (\nuv) &=\mathbf{0},\label{eq:PVP_}\\
\vv^t \xv(\nuv) &=1,\label{eq:norm}
\end{align}
\end{subequations}
are obtained with the \emph{bordered matrix} \cite{Andrew:1993}. 
This \emph{direct method} requires only the right eigenvector and preserves the global sparse structure of the matrix suitable for high performance linear solvers.
It is noteworthy that other approaches can be used to compute the derivatives. In particular, the approach from \cite{Nelson:1976} limits matrix changes or the method from \cite{Orchini:2021} may still be used for semisimple eigenvalues.

The bordered matrix is obtained by collecting the $\alphav$\textsuperscript{th} order derivative of the eigenvalue problem \eqref{eq:PVP_} together with the normalization condition \eqref{eq:norm}
\begin{equation*}
\begin{bmatrix}
\Lv & (\partial_\lambda \Lv) \xv \\
\vv^t & 0
\end{bmatrix}
\begin{pmatrix}
\partial^{\alphav} \xv\\ \partial^{\alphav}\lambda
\end{pmatrix}
= \begin{pmatrix}
\Fv_{\alphav}\\ 0
\end{pmatrix}.
\end{equation*}
The constant vector $\mathbf{v}$ is chosen proportional to $\partial_\lambda \Lv \xv$, though other normalizations are possible.  
The right-hand side (RHS) vector $\Fv_{\alphav}$ contains terms arising from previous order derivatives. 

Its general expression is cumbersome because of the nested dependency of $\nu$ in $\Lv (\lambda(\nuv), \nuv)$
though it can be explicitly obtained using Fa\'a Di Bruno formula. 
In the case of a single variable, examples of closed-form expressions of $\Fv_{\alphav}$ are available in \cite{Nennig:2020, Ghienne:2020}.
When $\Lv$ is a polynomial function of $\lambda$, which is the case in the examples treated in the present paper, the expression is amenable to a relatively easy treatment.
Let us consider a problem of the form (very general and used in the implementation)
\begin{equation}\label{eq:Li}
\Lv (\lambda(\nuv), \nuv) \xv (\nuv) = \left( \sum_{j=0}^T f_j(\lambda(\nuv)) \Kv_j(\nuv) \right) \xv(\nuv),
\end{equation}
where $f_j(\lambda)$ is a polynomial in $\lambda$ and $f_0(\lambda)=1$ (stiffness matrix). For instance, a generalized eigenvalue problem yields two matrices with constant and linear function of $\lambda$. The index of the eigenvalue and the eigenvector are dropped for clarity.
The successive partial derivative involves a product of either two terms $\Kv_i$, $\xv$  or three terms $\Kv_i$, $\xv$, $f_i$ and can be obtained thanks to the generalized Leibniz' rule Eq.~\eqref{eq:Leibniz_prod_mv}.

\begin{equation}\label{eq:Leibniz_code}
\left(\Kv_j \xv f_j\right)^{(\alphav)}=\sum _{|\kv^1 |=\alpha_1} \dots \sum _{|\kv^N |=\alpha_N} 
 {\binom{\alpha_1}{k^1_{1},k^1_2 ,k^1_3}} \dots {\binom{\alpha_N}{k^N_{1},k^N_{2} ,k^N_{3}}}
 \Kv_j^{(k^1_{1}, \dots, k^N_{1})} \xv^{(k^1_{2}, \dots, k^N_{2})} f_j^{(k^1_{3}, \dots, k^N_{3})}.
\end{equation}
This formula simplifies to
\begin{equation}\label{eq:Leibniz_codeK}
\sum _{|\kv^1 |=\alpha_1} \dots \sum _{|\kv^N |=\alpha_N} 
 {\binom{\alpha_1}{k^1_{1},k^1_2}} \dots {\binom{\alpha_N}{k^N_{1},k^N_{2}}}
 \Kv_j^{(k^1_{1}, \dots, k^N_{1})} \xv^{(k^1_{2}, \dots, k^N_{2})},
\end{equation}
 when $f_j$ is one (case of the stiffness matrix).

The partial derivatives of $\Kv_j$ are supposed to be available as well as the partial derivatives of $f_j(\lambda)$ with respect to $\lambda$. 
It is noteworthy that $f_j(\lambda)^{(\alphav)}$ does not depend explicitly on $\nuv$ but implicitly through the eigenvalue. This derivative is obtained through Faà di Bruno formula in the general case.

The $\Fv_{\alphav}$ is directly deduced from Eq.~\eqref{eq:Leibniz_code} and Eq.~\eqref{eq:Leibniz_codeK} once the left-hand side terms of Eq.~\eqref{eq:Bordered}, i.e. the products $\Lv \xv^{(\alphav)}$ and $(\partial_\lambda\Lv) \xv \lambda^{(\alphav)}$ are removed.
With the problem form \eqref{eq:Li}, it means that the last term of Faà Di Bruno formula, i.e. $(\partial_\lambda f_j) \lambda^{(\alphav)}$ has to be removed. %

The bordered matrix is factorized once for each eigenvalue $\lambda$. Then forward and backward substitutions are used to solve Eq.~\eqref{eq:Bordered} with multiple right-hand side vectors $\Fv_{\alphav}$. %

\section{Partial discriminant}\label{app:discriminant}

The previous approach introduced in \cite{Ghienne:2020, Nennig:2020} were based on the analytic function $h(\nu)$ to locate exceptional points. This function can be generalized through the discriminant concept. 
This analytic function of the matrix entries vanishes for all multiple roots of the characteristic polynomial and can be expressed from the eigenvalues
\begin{equation}
H(\nuv) = \prod_{1\le i<j\le M} (\lambda_i(\nu)-\lambda_j(\nu))^2.
\end{equation}
Similar to what was done for the characteristic polynomial, a truncated version of $H(\nuv)$, denoted $\mathcal{H}(\nuv)$, can be introduced when only some selected eigenvalues are retained in the subset $\Lambda$. 
This reads
\begin{equation}
\mathcal{H}(\nuv) = \prod_{p \in P} h_{p}(\nuv),
\end{equation}
where $h_{p}(\nuv)=(\lambda_i(\nu)-\lambda_j(\nu))^2$ for $i\neq j$. Here, we called $P$ the set of all possible different pairs of eigenvalues from the set of the selected eigenvalues $\Lambda$.
The number of pairs is $\#P=\binom{\abs{\Lambda}}{2}$. 
For instance if $\Lambda = \lbrace \lambda_1,\lambda_2,\lambda_3 \rbrace$, $P = \lbrace \pair{\lambda_1}{\lambda_2},\, \pair{\lambda_1}{\lambda_3},\, \pair{\lambda_2}{\lambda_3} \rbrace$.
As more eigenvalues are included in the regularization process, the radius of convergence is wider than if  single pairs of eigenvalues in $h$ were used, as in \cite{Ghienne:2020,Nennig:2020}.

A Taylor expansion of this function can be obtained from the successive derivatives $h_p$ depending themselves on the eigenvalues successive derivatives computed at $\nuv_0$ using the generalized Leibniz rules or, more efficiently, from the hierarchical expression given in Sec.~\ref{sec:hireachical}.

Once the truncated Taylor series 
\begin{equation}\label{eq:partial_disc}
\mathcal{T}_\mathcal{H}(\nuv) = \sum_{0\le\alphav\le \mathbf{D}} \frac{\partial^{\alphav}\mathcal{H}(\nuv_0)}{\alphav!} (\nuv_0 - \nuv)^{\alphav},
\end{equation}
has been constructed up to $\mathbf{D}=(D,\dots, D)$, EPs can be found among the roots. Such an approach is interesting since the eigenvalue is eliminated from the problem.
For the single parameter case, the polynomial is univariate in $\nu$ and the roots can be easily obtained from the companion matrix and the spurious ones can be filtered as in \cite{Nennig:2020} or using Eq.~\eqref{eq:sensitivity}. 
Nonetheless, the accuracy obtained with the partial discriminant was generally slightly worse than with Eq.~\eqref{eq:EP_system}.

When several parameters are used, the discriminant vanishes for all \EP{n} ($n\in [2, \dots N]$). Thus finding the roots numerically is harder than for the PCP because of the possible continuum of solutions like continuum of \EP{2} with two complex parameters.
The partial discriminant polynomial can also be used to quickly plot EPs structure using iso-value and to conveniently explore the parameter space.


\end{document}